\pdfoutput=1

\documentclass[onecolumn,11pt]{article}
\usepackage{ulem}  
\usepackage{graphicx}
\usepackage{xcolor}
\usepackage{comment}
\usepackage{multicol}
\definecolor{covercolor}{cmyk}{0.17,0.27,0.45,0.04}
\definecolor{textcolor}{cmyk}{0.44, 0.5, 0.68, 0.45}
\usepackage{kantlipsum}
\usepackage{hyperref}
\usepackage{xcolor}
\usepackage{indentfirst}
\usepackage{soul}
\newcommand{\f}{\begin{equation}}
\newcommand{\ff}{\end{equation}}
\usepackage{amsthm}
\usepackage{afterpage}
\begin{document}
	
\pagecolor{covercolor}\afterpage{\nopagecolor}

\thispagestyle{empty} 
\begin{figure}[h!]
\centering
\vspace*{-3.5cm} \hspace*{-2.6cm} 
\includegraphics[width=1.28\textwidth]{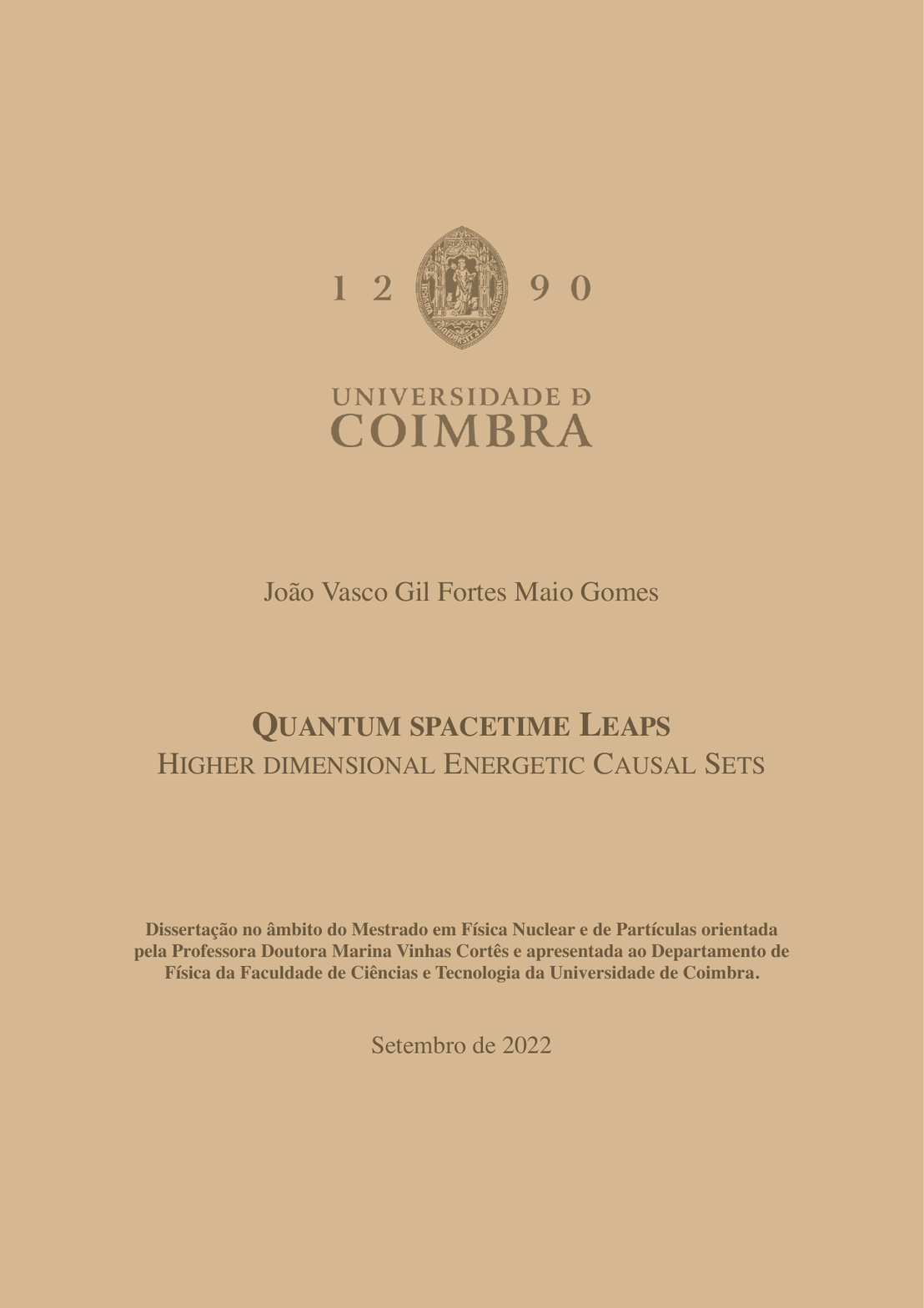}
\end{figure}
\clearpage 

~
\thispagestyle{empty}
\clearpage
  \begin{center} \textbf{Acknowledgements}\\~
  \end{center}
 
 I can think of a long list of people that helped me so far in becoming the person and the man I am today. To all those people, thank you, they know who they are. I would like however to specifically mention my family for all their support, in particular my parents.
 
 \paragraph{To my Father}
 
 Your presence in my life may have been shortened by the tragedy of existence, it was however enough to teach me a very important lesson. We never had those big father-son talks. You never taught me how to behave myself  and act properly as a human being, as a man, through big talks. You taught me that by setting an example through Your actions in everyday life. In doing so, not only did You teach me how to be a parent, a Father, but also, You taught me what a real leader looks like. Real leaders don't lead by imposing rules and their will on others. Real leaders lead by example and only those who lack the strength of character to be loyal to their principles need to lead  through oppression. People follow real leaders, while fake leaders need to make people follow in line. Take a look at today's world leaders, who become leaders to be served instead of serving... No wonder the world is in chaos...
 
 \paragraph{To my Mother}
 
 The list of things to thank You for would probably be bigger than the thesis itself, so to keep it simple, here's what I value most of everything I learned from You. You taught me to fight without giving up for what is right. You also taught me that some fights are just not worth our time while others might be worth our lives. I guess one of the secrets to life is knowing which are worth fighting and which are not. You also taught me that sometimes, no matter how lost someone might be, all they need is someone to believe in them even when and especially when they don't. That's basically what You did for me. So thank you for always believing in me Mother, I know it wasn't easy for You but rest assured, You can now rest. Now is up to me, thank You. \clearpage

~ \thispagestyle{empty}
\clearpage
  \begin{center} \textbf{Abstract}\\~
  \end{center}

     Why is the question of the passage of time technically so difficult to address, in the context of fundamental physics? Cort\^es and Smolin have proposed that the solution to the puzzle of the arrow of time, which has occupied the minds of intellectuals throughout the centuries, might lie at the intersection of quantum mechanics and gravitational theories, thereby forming a proposal for quantum gravity, along with a novel interpretation for the passing of time in foundational physics. They named their program Energetic Causal Sets (ECS). Since 2013, the solution existed only in the 1+1d case, and  the programme was halted due to the lack of models extending the 1+1d case to higher-dimensional manifolds. 
     
     In this thesis I successfully derived the mathematical framework for the 2+1d case. I then proceeded to implement my mathematical structure in a computational algorithm capable of withstanding the complex causal relationships between individual events in this higher dimensional extension. I also solved the obstacle of how to describe point-particle collisions in the real line. The metric of the causal structure from which the spacetime embedding emerges is the flat energy-momentum. This means that in 2+1d the set of solutions to the intersection of null rays has zero measure. In turn, this implies that point particles never interact, unless the initial conditions are specifically designed to do so, meaning the behaviour of the system is unnatural. 
     
     I found the same phase transition of the dynamics as Cort\^es and Smolin did in their seminal work. In the 1+1d case the phase transition was marked by the emergence of quasi-particles. The phase transition in 2+1d  takes the form of the emergence of a crystal, lattice-like structure. Further I was able to make progress beyond the initial goals of our project and find limit cycle-like behaviour in the 2+1d case. ECS in 1+1d models are typically captured by limit cycles in the context of random Boolean networks, described causal-time series of nodes in discrete dynamical systems (DDS). I was able not only to identify the limit cycles for the 2+1d model but also to describe how the proportion of deterministic versus indeterministic inputs to the algorithm affects the speed of attraction towards the basin of attraction (limit cycle). The phase transition of the dynamics that I obtained signals the evolution of my models into the time-symmetric regime. These are remarkable results for the progress of the ECS program, and validate a continued search for the recovery of the arrow of time in a foundational physics context.

\clearpage
\tableofcontents

\clearpage

\section{Introduction}

This chapter is devoted to a discussion of some of the key concepts involved in this work. The nature of the presented topics is quite profound and philosophical. As a masters student, and having had only one year, I cannot yet fully grasp how deep some of the concepts go. The titles in the first chapter might not be very appealing for today's physicist but bear with me. They will allow the reader to understand better where the ideas developed in this work come from. Furthermore let's not forget that back in the day physics, mathematics, and philosophy were all one discipline. The separation to three disciplines was inevitable; however, it is my personal belief that the separation has been too big, at least from physics. When we are trying to understand the world, to explain the fabric of reality and our experience, then leaving philosophical discussions out of the picture is, in my opinion, absurd. 

There is some evidence to support this, like the fact that we have been trying to solve a problem for a hundred years without success. For example at the birth of Quantum Mechanics there were lots of discussions involving its interpretation and the measurement problem. One of those interpretations, proposed by Wigner, was that a measurement is a process involving a conscious being. Even though not widely accepted, it was discussed thoroughly and as a consequence we still learn about it today. I think that unfortunately today in most groups of physicists such a proposal would be considered a nonsense from the start. I contend that once we try to go further than our current theories allows, discussions on philosophical grounds are necessary. 

\subsection{Why is the future different from the past?}

The most prominent feature of the world around us is the ever-present passing of time. Yet the equations supporting the pillars of modern physics are almost without exception symmetric in time. Those  equations admit a number of solutions in which time increases, and the same number of solutions in which time decreases. We discard half of the solutions by hand, reporting a solution back to the initial conditions of the evolving system under study. However, our time-symmetric-based physics gives origin to several arrows of time, which we describe here. Some of those constitute unsolved problems of contemporary physics: 
\begin{enumerate}
    \item One such example is the statistical mechanics limit of thermodynamics. The second law of thermodynamics imposes a clear distinction between past and future which is not reflected on the laws we take to be fundamental as quantum field theory and general relativity which are time symmetric. This means they are symmetric under reversal of the time coordinate which implies that for every solution there is a time-reversed solution in which past and future are interchanged. We never observe these last solutions and we justify this with highly time-asymmetric initial conditions. For example, we justify the second law of thermodynamics by saying that at the big bang the entropy of the Universe was extremely low.
    \item Another example is the electromagnetic arrow of time. Maxwell's equations admit both advanced and retarded solutions and yet, in nature,  we only ever observe the retarded solutions subset of those equations and never the advanced ones.
    \item the gravitational arrow of time: the universe contains no white holes, while black holes form to the future of the initial singularity. In 1979, in Ref.~\cite{Penrose1979}, Penrose proposed to solve this puzzle by asserting that the Weyl tensor vanish at the initial singularity, while reaching a maximum at a big crunch, or final singularity.
    \item The quantum arrow of time, which occurs when a system is measured and a quantum state is observed. The evolution breaks time symmetry, which in the quantum regime is referred to as a violation of the unitarity evolution of the wave packet.
\end{enumerate}

Each of these arrows in the time evolution of physical systems raises questions for the underlying theory. Most of these questions remain unanswered. Others have answers which themselves raise questions even harder to solve. It is fair to say that overall, there is no one proposal for the puzzle of the arrows of time that remains unchallenged by the physics community at large. An abridged discussion of these proposals and their challenges is outside the scope of this thesis.   

\subsection{Free will, indeterminism, and the passage of time}

This section allows us to give another perspective on ideas about Time developed in the work, which might help a reader who is being introduced to them for the first time to understand them better. These are not scientific truths, they are merely the discussion of ideas on philosophical grounds, something that I believe, as I just said, is necessary when we are trying to develop new and profound physical concepts of the world such as Time. 

A starting point leading to the choice of this thesis topic was the physics community's deterministic view of the world in the classical regime. Everything that has ever happened and will ever happen is pre-determined by an initial set of conditions. Not only does this lead to quite a tedious and monotonous Universe, but also to a somewhat authoritarian one.

We are part of this Universe, and so are our lives and everything that happens in those in the future. Is that pre-determined as well? Is there no `Novelty' in our future? How can a deterministic view of the reality accommodate `Free Will', in which humans, mostly, wish to be a feature of our existence? Surely our belief in `Free Will' does not mean nature attends to our request. As scientists and students of nature we must come to our field devoid of prejudices, inner conviction, or personal taste. So firstly we must address why should the existence or agency of `Free Will'  be a logical, and necessary, truth.
For, as Einstein notoriously expressed,
\begin{quote}
{\it Everything should be made as simple as possible, but not simpler.}\\
\hspace*{11cm} Albert Einstein
\end{quote}

\subsubsection{Societal self-organization}
Let us then assume the future is pre-determined as in the block universe view~\cite{marinablock}. We can think of the ethical questions that a pre-determined future entails. Any one member of society carrying out actions of offense or violations of civil law cannot, technically and consistently be hold accountable, nor are liable for their actions. After all, according to the block universe view, such society members have been conditioned since the beginning of the universe to undertake such unlawful actions. A civil legal system is the product of reflected intellectual thinking on how best to find the set of principles by which each society can govern itself. If an intellectually-organized civil regime carries deep inconsistencies in its foundations, it cannot play a role as the primary source of a trustworthy, self-consistent, and referable core of principles supporting persistent societal interaction.

In the aforementioned societal organizational structure we can raise two questions:
\begin{enumerate}
    \item why should any society members feel responsible or regret for their unlawful actions?;
    \item how can such a system give rise to persistent social interaction?
\end{enumerate}

Theoretical physics is the study of reality, and as students of reality we aim at making sense of the the world around us. Theoretical physicists are often viewed as upholders of the ultimate laws of nature which support the existence of the universe. Such responsibility bestowed upon our community merits therefore a self-consistent reflection of the world views we hold, that we then pass on to society.

\subsubsection{Accommodating ``Free Will''}
Is there any alternative to the block universe view of a pre-determined future awaiting us?
I have then found precisely this in a paper by Nicolas Gisin \cite{freewill} which we refer to in Section~\ref{sec:freewill}, the next section. But if we take Free Will as fundamental, that immediately means that the future is open and the world can't be deterministic, for if Free Will is the ability to make choices, certain aspects about the future can't yet be determined. And that also suggests that time must really pass. What this means is, if we agree there are aspects of the future that might not be determined by an initial set of conditions, then there is a clear distinction between past and future and as physicists we aim at the explicit inclusion of this distinction in the equations governing the evolution of the universe. We hope that this thesis will pose a good example of how a an abstract discussion can be translated onto concrete mathematical expression for the evolution of the system being studied. If this translation hadn't been found, nor others like this, we could not hope to formulate new physics under the idea that maybe the true fundamental laws of Nature are time asymmetric. The traditional view in physics is that the variable `time is emergent in a regime of quantum gravity, and results in the observed asymmetry to be a consequence of improbable choice of initial conditions\footnote{Intelligent design? Multiverse?}. Here we will propose the alternative view: the variable time is governed by fundamentally irreversible dynamics, and is therefore present before the emergence of spacetime in searches for quantum gravity.

\subsubsection{The passage of time}\label{sec:freewill}
Free Will, Gisin argues, is a pre-requisite for rational thought since we depend on the ability to choose which arguments to take as true and which arguments to discard. It is therefore not trivial to prove its non-existence, since that proof requires rational reasoning which in turn requires ``Free Will''\footnote{I suggest to the reader the possible reflection: if we do not believe in the existence of ``Fee Will'', we are using our own ``Free Will'' to choose not to believe in its existence.}. If we choose to proceed under the assumption that the current equations in physics do not suffice to determine the entire future evolution of the cosmos, including human existence, then we might be helped by the search of a theory in which the passage of time features at the foundations of physics. The question then is which `time' variable passes? If we take the `time' variable describing harmonic oscillators, which is an example of a time-symmetric system in the zero friction limit, we won't go to far since this time only suffices to describe the unfolding of a symmetry, where nothing new ever happens. This concept of time is to poor to describe our experience.
In order to address the passage of time it is useful to mention two different concepts of `time'
\begin{enumerate}
    \item creative time or Heraclitus time and,
    \item geometric, or Parmenides, time where nothing new happens.
\end{enumerate}\label{heraclitus}
How do we then introduce indeterminism or novel evolution?
One the possibilities lies in the belief in random number generators, which can account for genuine non-deterministic evolution. We are used to think of probability and randomness in a statistical way, meaning we think of it as a property of a set of events and not of as a property of a singular event. But what characterizes a single random event? How is it created? A random event is created when two causal chains meet. What characterizes the event as random, quantum mechanically speaking, is the lack of knowledge one might have about the causal chains in question. From the moment we have that knowledge, the random event becomes predictable, although not totally predictable for the following reasons.

One can argue, and rightfully so, that having a complete knowledge of the present would enable the event should be totally predictable. But we know that quantum uncertainty prevents a full knowledge at any instant in time of all the variables that constitute reality. This translates here to the fact that one cannot, simultaneously, possess knowledge on all causal chains of events happening in the Universe. This happens because an observation is local, something we will discuss in Section~\ref{sec:becoming}. It is true that the event happens, let's say at the junction of two causal chains. However there will always be at least one causal chain that we don't have possess  knowledge of--or no knowledge at all-- and that prevents us from predicting with absolute certainty the next random event. 

In this case the best description of the future event is given by its probability, the chance that that event might come into $being$\footnote{This is the prevailing interpretation of Quantum Mechanics.} The \textit{happening} of such an event marks an abrupt distinction between what happened until that point in time and that which happens from that point forth. In other words there is a Universe before that point in time which is distinguishable from the Universe from that point on, something new was created. Clearly a concept of time that merely describes the unfolding of a symmetry can't allow for the inclusion of something like an explicit mechanism of {\it happening} and its featuring explicitly in our mathematical description of reality. This is the action of creative, Heraclitus, time, mentioned above. And this is also why we will use (machine) randomness in our work to produce a fundamental irreversibility.

If we hypothesise that no future event is certain, the future is still open and events always have a certain probability to happen. Heraclitus time leaves open the chance for emergent phenomena at late time and large scales to occur. Even though this explains many phenomena and their irreversibility, it raises the question, on the Cosmological scale, of what chooses these improbable initial conditions and why they should be time asymmetrical. Is it just an accident, which is today's common belief, or a hint of a new fundamental principle that makes past and future truly different? Roger Penrose \cite{Penrose1979} already made the radical proposal a few decades ago that maybe the fundamental laws are time asymmetric and that on larger scales these could be approximated by emergent time-reversible laws with the imposition of time irreversible initial conditions. 

Given these properties of creative time, which we will simulate using randomness, it's only natural to make it, as we shall see, the total order in the set, corresponding to the birth order of the events (this as nothing to do with a Minkowski birth time order).

\subsection{Becoming}\label{sec:becoming}

Creative time, mentioned in Section~\ref{heraclitus}, describes much better our experience of the passage of time, as the unfolding of unpredictable, in the sense described, events that come into \textit{being} and create a clear distinction between past and future.  
But what is the sense of the passing of time we all share? That is what is called becoming \cite{becoming,becoming1,leesmolin}. Becoming is nothing more than the $happening$ of events, when events come into $being$. It is the precondition for an event to exist in spacetime, in our Universe. Becoming is what gives us our sense of Now and flow of time. But how should we characterize the Now? I mean, should the Now be Universal or should it be local? Should we therefore try to make sense of becoming in a local or global manner? 

``Now'' is depicted in traditional view as a spacelike hypersurface, and in the same view the flow of time is an illusion, time is just a succession of these global Nows stacked one after the other to form spacetime. This is, after all, the Block Universe picture summarized in Ref~\cite{marinablock}. So becoming should, according to this point of view, be global. However simultaneity of spacelike events has to be established from definition \cite{einstein} and spacelike events have no influence on the observations that give us the sense that time is flowing since these are local \cite{becoming}. We cannot observe directly distant points in space from our location since we need information from those points to get to us and this information doesn't travel to us at an infinite speed. This is to my understanding the essence of Relativity and I don't think it gets enough emphasis in general, so let's talk a bit more about it. 

Relative simultaneity describes how two observers moving relative to each other might disagree about whether or not two events are simultaneous. This led me to think of the following example. \begin{enumerate}
    \item Imagine two observers in one dimension, observer (1) in the middle of a platform and stationary relative to it and observer (2) cruising by. Let's say observer (2) sees observer (1) traveling, with the platform, to the right at speed $v$. 
    \item There are two light bulbs, one at each end of the platform and at a certain instant of time the light bulbs are turned on.
    \item Let's say the observer (1) at rest on the platform records the light arriving from each bulb at the same time. What will the moving observer (2) see?
\end{enumerate}

Einstein realized that since the speed of light is the same for both observers, $c$ and since observer (2) perceives observer (1) moving towards the ray incoming from the right bulb and away from the ray incoming from the left bulb, then the light must have departed the left bulb first, as measured by observer (2). This is because observer (1) says both rays reached them at the same time so, by the principle of relativity, observer (2) also needs to see both rays of light getting to observer (1) at the same time\footnote{The local time of observer (2) naturally.}. Therefore, to compensate for the fact that observer (1) is moving away from one ray and towards the other, the events, the turning on of the light bulbs, won't be simultaneous to observer (2).

However this led me to the following reflection. I clearly set up the experiment with the observer at rest in middle of the platform to simplify things. But now I can take the two observers, both at rest relative to the platform, at different positions and they will also disagree about whether the light rays reach them at the same time. Let's say we have one of them in the middle, the other at the position of one of the light bulbs and the one on the middle sees again the light bulbs being turned on at the same time. The other observer will say they received the light from the bulb at their position first. So they don't seem to agree as well in this case.  This can be confusing since it doesn't involve any relative motion and the only reason the second observer doesn't see both light bulbs being turned on at the same time is the fact that they are at a different distance from the first observer. The last paragraph might seem confusing and incorrect but is in fact related to Einstein's critique of absolute simultaneity \cite{einstein} which is not just related to relatively moving observers. These two observers disagree because they can only have access to the temporal relations of events in their own immediate surroundings.

Einstein realized that the problem here is that these observers are recording events that are far from their clocks and their clocks can only give an accurate temporal order of events happening at their own spatial positions. In order for both of them to get an accurate order of events in this case, they need to define a common time between different places, but this has to be established from definition. So Einstein says that two clocks will be synchronous if the time it takes light to travel between them is the same in both directions. Note that this is the same as saying that the speed of light is direction independent and we already saw in the first example that if there is relative motion between the clocks, the speed of light becomes direction dependent. This really defines simultaneity, for it is the same as saying at the clock where the light started the journey, the event halfway between the emission of the light and its return is simultaneous with its reflection at the other clock.

Let's now see how the two observers proceed, after having agreed to this definition. Knowing their clocks are synchronous with clocks at the light bulbs and knowing the distance they are from each light bulb, they can now both measure, by subtracting the time it took light to travel to them from each bulb, that in fact, the two light bulbs were turned on at the same $common$ time. This seems like a game of words but it isn't, it seems obvious that they should both take the time it took light to travel to them, what was not so obvious was that simultaneity had to be established from definition. The Devil is in the details and that's why it took an Albert Einstein to figure this out. They will still observe different things but now they measure the same thing. To observe something is to merely record it happening. To measure something implies a treatment of the mere observation that will require, in general, established rules. The rule that needs to be defined here for the observers to measure the same result is what it means for two clocks to be synchronous, in other words, how do we define simultaneity for distant events.

For clarity let's summarise the argument above. We started by talking about observers moving relative to each other and disagreeing whether or not two events were simultaneous. That was because it is my understanding that, generally speaking, this question about simultaneity is perceived as being only related to relative motion when in fact it isn't. So I wanted to make a clear distinction between both scenarios, with and without relative motion. After that I tried to make it clear that if two distant observers want to talk about a common time between them they first need to establish what that means by definition. This is to put emphasis on the fact that observations are local and this is the main point because remember, I was trying to answer the question of whether or not we should consider becoming to be a global or a local phenomenon. Observations are local, in particular those that give us a sense of flow of time, of $becoming$. Therefore while relativity itself does not support the need to see time as a succession of cosmic nows, it naturally doesn't contradict it either and I am not saying that. Furthermore, this problem with simultaneity only applies to spacelike events relative to the observer (represented by the apex on their spacetime diagram). But these events cannot influence or be influenced by the observer. They therefore cannot influence one's perception of $becoming$ which diminishes even more the view of now as a spacetime hyperplane extending through all space linking all simultaneous events at a certain instant. In fact it would seem more appropriate to treat time as we experience it and the sense of now as a local thing. So maybe we should try to make sense of $becoming$ as a local thing as well.

There are other physical reasons to support the fact that describing the flow of time, and therefore $becoming$, as a succession of global nows is not the best approach \cite{becoming}. First it is possible, following Reichenbach's critique of the definition of simultaneity \cite{becoming,reichenbach}, to come up with, different definitions of simultaneity that lead to physical and empirically equal spacetimes, although only one of these, the relation of Minkowski orthogonality with respect to each observer, which is Einstein's definition, is built into the metric of Minkowski spacetime \cite{becoming,reichenbach}. The idea is that you can alter the definition without changing the spacelike nature of the events. And since these events can't influence or being influenced by an observer, their observations of the Universe are not influenced by which spacelike events are defined to be simultaneous. To make this more clear take a look at Reichenbach's formula for a definition of simultaneity \cite{reichenbach}:

\f
t_1=t_0+\epsilon(t_2-t_0)\,.
\ff

Take $t_0$ to be the time light leaves clock A, $t_1$ the time it is reflected at clock B and $t_2$ the time of its arrival back at A. All these instants correspond to the time at A and that's exactly why we need a definition to go further and talk about a common time with B. Einstein's definition takes $\epsilon=1/2$ (so it is really saying that the reflection at B and the event at A at half of the journey are simultaneous) but any value between $]0,1[$ doesn't change the spacelike nature of the events, we just take another reference as the definition, like taking a different reference for what a kilogram is. This naturally leads to different definitions of now hyperplanes and even if we want to take Einstein's definition as the fundamental one, this still gives rise to more than one definition of now hyperplanes (one of each observer\footnote{Even at constant speeds, different observers will have different trajectories in spacetime leading to different orthogonal hyperplanes of now.}) and doesn't therefore lead to a consistent definition of a global now, unless of course we are willing to say that there is a privileged observer in the Universe. It doesn't therefore lead to a consistent way of defining a global now. 

Moving on to general relativity things get even worse since there are solutions to Einstein's equations that can't be sliced into spacelike hypersurfaces like in G\"odel Universes \cite{becoming,godel} where closed timelike worldlines exist.\footnote{It's impossible to define a succession of global nows with closed timelike worldlines.} Of course these Universes are purely theoretical as far as we know, so let us for example take the Robertson-Walker solutions. These describe an isotropic and homogeneous Universe, which our Universe is not so this doesn't even take into account small scale phenomena. These solutions allow for the slicing of spacetime into spacelike hypersurfaces. One could argue \cite{becoming} that by averaging over regions of spacetime, one gets a really good approximation to our Universe and can therefore define global instants of time. However the way one would come up with these global nows, the spacelike hypersurfaces, would be dependent on the averaging process and the size of the spacetime regions the average was being made on. This again does not lead to a consistent definition of what is a global now.

In conclusion those observations that give us a sense of flow of time and $becoming$ are independent of a global now-slicing. This, together with the fact that we can't uniquely define a global now-slice, means that we should try to make sense of becoming as a local phenomenon and look at spacetime not as a linear ordered set of events but as partial ordered set of events. In fact Energetic Causal Sets are inspired by the idea of $becoming$ \cite{paper4}; when creative time is taking place, interactions are taking place in pre-spacetime with energy-momenta being prescribed in this regime and only then spacetime emerges allowing events to come into $being$. As a final remark, note that the fact that we should look at spacetime as a partially-ordered set doesn't contradict a total causal order in the pre-spacetime regime. In fact if this order, created by creative time, imposes the distinction between past and future at the most fundamental level, then it should be a total causal order in the pre-spacetime regime. Given the local nature we want to attribute to becoming, the law responsible for evolving creative time should be local.\footnote{This will be the objective of future work in collaboration with Rafael Sorkin}


\newpage
\section{State of the Art}

In this thesis only the classical version of energetic causal sets is extended to 2+1d, therefore it's quantum counterpart \cite{paper2} will not be discussed. Also the work developed so far only applies to flat spacetimes. 

\subsection{Energetic Causal Sets}\label{ecsintro}

Energetic Causal Sets \cite{paper1,paper2} is a recently-introduced topic seeking to create a system that allows one to study, among other things, the ideas described in the introduction. The work is motivated by the idea that the fundamental laws of physics should be time asymmetric, not time symmetric, and that causality, resulting from the irreversibility of time, is the fundamental principle that governs physical processes. 

Even though the name is suggestive, these are fundamentally different from the usual causal sets \cite{causalsets,csfd,csfd1,csfaydowker1}. The main difference is that energetic causal sets allow exchanges of energy and momentum, within the set, that is, in pre-spacetime. This is accomplished by taking energy-momentum as fundamental properties of the events, so in effect they live in energy-momentum space with an assigned Minkowski metric (since we are only dealing with flat spacetimes here). The biggest consequence of this is that one can now have built in dynamics in the set by applying relativistic energy-momentum conservation laws at each event instead of having them to be emergent. There are four principles that underlie the foundation of energetic causal sets \cite{paper1,lee1}:
\begin{itemize}
	
	\item Time is a fundamental quantity, an elementary process by which new events are created out of present ones. Causality is a result of this process and its irreversibility. This is the activity of Heraclitus time and is the total causal order on the set.
	
	\item Time (the creative time, the time of the first principle) is unidirectional in the sense that we cannot use the true fundamental laws of the Universe to reconstruct the past out of a complete knowledge of the present.
	
	\item All spacetime properties have a dynamical origin and spacetime itself is to emerge from the dynamics, with the time being the Parmenides time. This is responsible for the partial order induced by the metric, flat in our case.
	
	\item Energy-momentum is a fundamental property of each event and through laws of conservation dynamics arises. This also gives birth to a dynamical partial order on the set that basically keeps track of the transfers of energy-momenta which are controlled by energy-momentum conservation laws.  
	
\end{itemize}

The first two principles refer to the nature of time in the model while the are two contemplate the dynamics. In view of these principles there are three causal structures at play here. The first, created by the agency of Heraclitus time, that brings novelty into the Universe and allows the possibility for $becoming$, is a total order since it is the order by which events are created. Then we have two partial orders, a dynamical partial order that prescribes the flow of energy momentum between causally related events and an emergent causal partial order that reflects the causal network in an embedding in spacetime. Both the total order and the partial dynamical order are microscopic orders, they belong to the set, while the emergent causal order is a macroscopic order induced by the metric of the emegent spacetime. 

The third principle also makes it so that every event is unique. Since all spacetime properties of an event have a dynamical origin they must arise from the (dynamical) relationship of this event to all other events. This has the important consequence that each event is distinguishable from all others by its relational properties. 

The idea to describe spacetime at the fundamental level as a causal set comes from the fact that knowing the causal relations between events i.e. the causal network, makes it possible to reconstruct the metric and therefore spacetime. So one can imagine that fundamentally spacetime consists of a discrete set of events together with their causal relations from which spacetime is to emerge as a bulk average of these relations over the set. From this point of view there are two partial causal orders in causal sets, the microscopic causal order of the set (that reflects the causal network within the set) and an emergent causal order that appears with the embedding of the set in a suitable spacetime, a macroscopic causal order (the one induced by the metric). In typical causal sets these two causal orders always agree. This means the arrows of time defined by the two causal structures are always equally aligned. This will not always be the case in energetic causal sets, the emergent causal order doesn't need to preserve the other two, a phenomenon called discausality, or disordered causality \cite{paper4}.

The way events are generated is also fundamentally different. In causal sets two events are picked randomly to generate a new event \cite{csfaydowker}. In energetic causal sets, since we now have built in dynamics in the set we can have a rule to pick which, out of the set of available events, will be the parents. Events need to be unique for this to make sense and indeed they are. Within a discrete causal structure the only way an event can be distinguished from another is by its causal past. So in order to create a rule that picks which events are going to generate the next event a measure for the causal past of each event will be needed.

Another difference is that while in causal sets all events are always able to be parents, in energetic causal sets, once an event gives birth to two other events it can no longer interact. This has the implication that unlike in causal sets there is a big difference between past and present built into the model. We can now call the union of all exhausted events the past and the union of all available events the present. Given that an event with one child is still able to interact one more time it belongs to the present so we can have a scenario where causally related events can be in the present. Such a present is called a thick present for this reason. This is what produces discausality in the set. This is considered a form of retrocausality also with implications for quantum foundations in particular a realist completion of quantum mechanics \cite{paper3, paper4} and the two-state vector formalism \cite{tsvf}. 

With regards to the rule it needs to captivate the discussed ideas about time. It needs to be irreversible and local and given that we can only distinguish events by their pasts the rule will have to pick the parents based on their pasts. We don't want dynamics to be cumbersome, so a rule based on extremising the differences in the pasts will suffice. In particular we want to minimize this difference to make the rule local. Random inputs are introduced at the generation of a new event in order to impose irreversibility. Together with the fact that we have a clear distinction between past and present we are able to build into the model a fundamental directionality of time described by the first principle in the set. The future does not yet exist since we can't form the union of events that haven't yet been created, its a null set. And given that there is a random input in the code, at best, only the probabilities of occurrence of a future event could be predicted.

The energetic causal set model enables the study of how an effective time reversible dynamics can emerge at late times and large scales from time irreversible laws and that is its main purpose. The time symmetric phase of the model is tied to the capture of the system by limit cycles in discrete dynamical systems (DDS). This phase is time symmetric in the sense that a system reduced to it's limit cycles is time symmetric \cite{paper4}. 
Discrete dynamical systems are a large class of dynamical systems described by discrete and deterministic evolution rules, in which each state a unique successor chosen from a set of possible states \cite{AW}. They have the following properties
\begin{enumerate}

\item{}There are a large, but finite number of discrete states $S= \{ I,J,K,L, \ldots \}.$  
\footnote{Or in the quantum version, a finite dimensional-state space.}
\item{}The system is governed by a discrete and deterministic evolution rule. Thus each state, $I$, has a unique successor:
\f 
I \rightarrow J
\ff
\item{} Evolution is surjective, i.e.\ there is no rule constraining the number of precedents a state $I$ may have.  In particular, there may be states with several precedents.
\end{enumerate}

Generic dynamical rules of this kind are irreversible, and the underlying dynamics is indeed discrete and time-irreversible. But in the long term they asymptote to time reversible dynamics and exhibit such properties. This behaviour is captured in the form of a limit cycle, and it is well known that the limit cycle attractor is ubiquitous in DDS \cite{AW}. 

\begin{figure}[t!]
\centering
\hspace*{-0.5cm} 
\includegraphics[width= 1.1\textwidth]{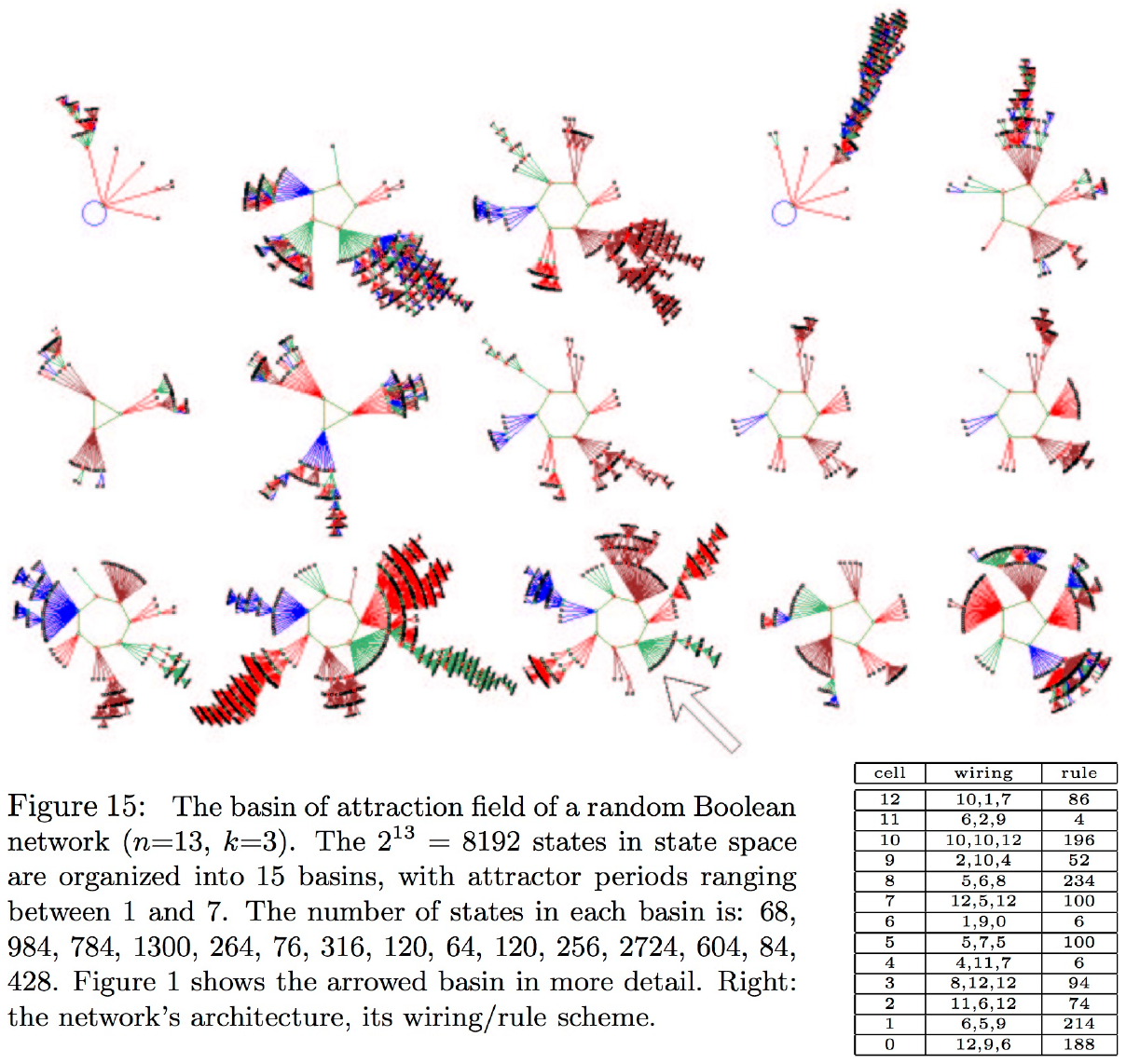}
\caption{The complete state space of a discrete deterministic dynamical system, showing several limit cycles and their basins of attraction. This behaviour of time-irreversible dynamics reaching time-reversible behaviour was identified with the phase transition observed in energetic causal sets in Ref.~\cite{limitcycle}. Image courtesy of Andy Wuensche from Ref.~\cite{AW}.}
\label{figureDDS} 
\end{figure}

Generically, limit cycles are basins of attraction in the dynamics of random Boolean networks. A pictorial example of such basins of attraction in given in Figure~\ref{figureDDS}.
%
%
The limit cycle phase is symmetric under reversal of the limit cycle, because, once in the cycle, every state has a unique antecedent and a unique descendent. Our objective is therefore to observe this phase transition in 2d like it was observed in 1d in Ref.~\cite{limitcycle}.

\subsection{Simulations in 1d}\label{1dsim}

Let's now talk about the simulations of a classical 1d model. Each event is composed of a pair of incoming photons and a pair of outgoing photons. In each pair one photon is left moving and the other right moving. We will need to make our emergent space compact with boundary conditions being applied otherwise, photons cross each other once and no more interactions occur. That will require momenta to come in integral units of a fundamental quantity \cite{paper1}. Photons are used since their dynamics are simple and they will trivially satisfy the constraints to be imposed on the transfers of energy-momentum between causally related events. The system will be simplified as much as possible in order to obtain a simple emergence of spacetime which we will replicate in the 2+1d case as well. The important thing is that whatever choices we make to simplify the the computations and the code we must stick with them consistently. If we denote $p^I_{aK}$ the four momentum incoming to event $I$, coming from event $K$ and $q^L_{aI}$ the momentum outgoing from event $I$ to event $L$, with $a$ the spacetime index then conservation of momentum means:
\f
\it{P^I_a}=\sum_K{p^I_{aK}}-\sum_L{q^L_{aI}}=0\,.
\label{pconservation}
\ff
where the sum over $K$ is for all past events connected to event $I$ and the sum over $L$ for all future events event $I$ is connected to. For simplicity we are adopting a static Minkowski energy-momentum metric, so there are no redshifts. This means that
\f
p^I_{aK}-q^I_{aK}=0\,.
\label{noredshift}
\ff
Note that if we wish to include expansion of spacetime in a cosmological setting, this is readily achieved by the introduction of an expansion/contracting function on the right-hand side of Eq.~\ref{noredshift}. Finally and since we are dealing with photons, energy momenta are constrained by the usual relations:
\f
\eta^{ab}p^I_{aK}p^I_{bK}=0,  \;\;  \eta^{ab}q^I_{aK}q^I_{bK}=0\,.
\label{eprelations}
\ff
Remember there is no spacetime and events live in energy-momentum space, so $\eta^{ab}$ is the metric in this space which naturally is taken to be the Minkowski metric. The system is set up in such a way that all these equations are trivially satisfied. Photons are chosen for this reason, since when they interact they simply cross each other without altering their momenta. 

Now is time to think about a measure of the causal past of each event. It has to be simple enough to be computationally feasible and still be able to capture the complexity of the causal past. As more events are created the causal pasts get more and more complex. The way around this is to realize that all that complexity can be captured by spacetime coordinates of an emergent spacetime. Another reason why we want the system as simple as possible is to ensure this emergence always happens. Armed with spacetime coordinates one can easily come up with a simple measure for each past:
\f
past_I^2=\frac{1}{N}\sum_J(-t_J^2+x_J^2)\,.
\label{1dpast}
\ff
Here $x_J$ takes values in compact space and $N$ is the number of events $J$ in the past of event $I$. This clearly makes each past unique (remember that was the goal, to be able to distinguish an event from all others) and is easy to compute. 

You can see this measure as sort of an average of spacetime positions. In order to calculate Eq.~\ref{1dpast} we will need to keep track of the causal past of each event. To that end events are stored as families, one family per initial event, and the family an event gets stored in is the family of the right moving parent.\footnote{The left-moving parent could have also been chosen. The important thing is to consistently choose the same parent.} One can argue that this leads to a loss of information of the causal structure but this is of no consequence for the results since it doesn't change the dynamics and has the benefit of simplifying the model. To chose the parents we simply take the difference between all the pasts of the events in the thick present:
\f
D_{IK}=|past_I^2-past_K^2|\,.
\label{pastdiff1d}
\ff
where the modulus is used instead of a square root to reduce the computational demand. The interacting pair is the one for which the value in Eq.~\ref{pastdiff1d} takes a minimum. 

The last thing we need to take into account in this model is the fact that if we allow photons to interact every time they cross, due to the boundary conditions, we will end up with a system equivalent to one of harmonic oscillators where each photon simply moves back and forward. Such a system is of no interest for our purposes. Given the parametric equations of the spacetime trajectory of each photon one can calculate all future interactions for all pairs given the initial conditions. Also, since we are in 1d all opposite moving photons (which are the only ones able to interact) cross each other before crossing the boundary. So, we can decide how many times photons have to cross the boundary before interacting and this is also how randomness is introduced into the model, by randomly selecting how many times the boundary is crossed, how many windings happen, before the selected pair interacts.

The simulation starts by randomly generating initial positions for the events at $t_{Minkowski}=0$. At each step two events are selected to generate a new event. Then the thick present is updated together with the family that will inherit the new event. This process in then repeated for the number of events, $N_{{\rm events}}$, we wish to create. \newpage

\section{Simulations of Energetic Causal Sets in 2+1d}

In this section we will describe our simulations, which are carried out using the \texttt{Mathematica} environment. Everything will follow as closely as possible to the 1d case. We have three programs corresponding each to three different selection rules. We have a fully-random code where parents are selected randomly, a fully-deterministic one which uses the rule of the 1d case adapted to two spatial dimensions,
\f
past_I^2=\frac{1}{N}\sum_J \left(-t_J^2+x_J^2+y_J^2 \right)\,,
\label{2dpast}
\ff
and a third program which combines the other two, meaning that sometimes the parents are selected randomly and other times using the rule, with the amount of randomness being controlled by a variable. A version of this code is presented in the Appendix. The structure of all the programs is basically the same with differences arising mainly in the selection of the parents. With that in mind we will first describe the overall structure of the programs and in further subsections we will explore in more detail what differs in each one.

We are envisioning a system of particles and anti-particles. To make the code as simple as possible, these are chosen again to behave like photons that simply cross each other without altering their momenta. Only particles and anti-particles can interact, there are no interactions between particles or between anti-particles. Since they are photon-like\footnote{From now on we will be refering just as photons.} one could ask, why make this distinction? This is to ensure that all families\footnote{We will store events in families in order to keep track of each event's causal past.} remain present and available for interaction for the entire duration of the run. In the 1d case what granted this was the fact that the family that would inherit a new event would be the family of the right travelling photon. Similarly only photons moving in opposite directions were able to interact. In order to replicate this in the 2d case we make a distinction between particles and anti-particles so we can have the new event always belonging to the family of the particle. We tried using the photon that was travelling closer to the $x$-axis and other variants but none seemed to work, we never got anything close to what could be called an even distribution of events among the families, at least for the program where parents are selected randomly, which meat there were families that were no longer available for interaction. This however, solved the issue. We will come later to this problem, for now keep in mind that we have a system of thick particles and anti-particles that don't alter their momentum on interaction, they just go straight through each other.

In order to enforce that only particles and anti-particles interact, we will need two tables to keep record of the available events for interaction, one for the particles and one for the anti-particles. These tables will represent our thick present. So we start by creating these tables, each with the selected number of families.

We want to do this in such a way that at global Minkowski time $t_{Minkowski}=0$ events are given random positions in our compact space. Particles are given random initial directions with the corresponding anti-particle moving in the opposite direction.\footnote{Naturally this only happens at $t_{\rm Minkowski}=0$ after rays and anti-rays coming from different events start to mix this will not be the case.} This corresponds to our set of random initial conditions. So we have $0$ total momentum in the system and Eqs.~\ref{pconservation} through \ref{eprelations} are trivially satisfied.

Once the initial events, corresponding again to the number of families (or causal pasts) we will keep track of, we use a For cycle to generate as many new events as we want. The first step is to select which events will be the parents. We will come back to this later since this is program specific. Two events interact twice and interaction moves on to another part just like in the 1d case. Remember, from each event leaves a ray and anti-ray so each event can only interact twice. We envision our particles and anti-particles as 2d discs of radius $R$, this is needed for interactions occur. Rays coming out of the same event are not allowed to interact in order to increase the diversity in the ``genetic pool". 

We use the biological term ``genetic pool" to represent the existing diversity in the dynamical evolution of the 2+1d system. This has an analog in evolutionary theory named the inbreeding problem of reproductive systems. The genetic pool of living systems needs to have enough diversity for sustainability of the system. In the 2+1d case there needs to be enough diversity so that the looked for phase transition to the crystal structure, as we will show below for example in figure~\ref{det2fam}, doesn't occur promptly when the simulation starts. Once we have our events, one from one table and the other from the other table, we immediately know the particle and anti-particle that we want to compute the interaction for. Every time two rays cross meaning, the distance between the centers of their discs becomes smaller than a diameter, we have an interaction. We compute this by using the parametric equations of the spacetime trajectory of the center of each ray, something we will see in detail once we discuss the computation of the intersections. Finally we have to update our thick present removing the old rays and including the new ones with new updated spacetime positions.  We also need a table to record the spacetime positions and respective families of all the events so that in the end we can plot them in their respective families.

This was just a summary, an overview of what the programs need, now we will discuss everything in more detail starting with tables and their information, which is necessary to build the programs.

\subsection{Parameters}

Before moving on to a description of the code's body we will talk about out the parameters involved that will control the runs. First we are working in compact space, so it would be good to have variables to define our 2d square/rectangle, $L$ and $W$, for length and width which are set equal to each other. For simplicity we will make our compact space a square since this doesn't alter the results.

Since particles have a 2d measure we also need to define the radius of the circles representing the particles and anti-particles. The radius itself hasn't much meaning to the results, only how big or small it is in proportion to our space, so we define the fraction of the radius and the side of the square as the variable to control the size of our particles and anti-particles. We called this variable the cross-section, $\bar{\sigma}$.

We don't want to wait an arbitrarily long time for interactions to happen so we need to define a variable, called $T_{{\rm max}}$. This is the maximum time we will be willing to wait for an interaction to occur and therefore for how long we will compute the coordinates. 

Finally and since we are using the parametric equations of the centers of the circles to find the interactions, we will need to define a time step $dt$ to iterate through time. These variables can be included in the set of initial conditions since, as we will see, they influence the behaviour of the system. 
 
The parameters we just mentioned are dependent on one another. For example, it would be wise to write $T_{{\rm max}}$ in terms of our spatial dimensions, since given that since the speed of the rays is 1, how many times the length L\footnote{We are working with a square so we just need to mention one of the dimensions.} we are willing to wait translates directly into how many times rays get to cross the boundaries before interacting. The variable $dt$ is also not independent of the radius, since if it is too big compared to the radius we will miss interactions. Make it to small and the program will become to heavy, so what is too big and too small? Well we want a good compromise between precision and computational time. From all the runs we made we concluded that in general, a good rule of thumb it to take $dt=\bar{\sigma}$. We will be coming back to these relations through out the text.

\subsection{Tables}

\subsubsection{``livev1'', ``livev2'' and ``allevent''}

These tables will, like we said, represent our thick present, ``livev1'' with the particles and ``livev2'' with the anti-particles ``livev'' for live events. Each row of each table pertains to a ray or anti-ray from a certain event and each row has all the information we will need. First, it has the number of the event. This number represents the order by which events are created and corresponds to the total causal order in the set. This order will be reversible or irreversible depending on the way the parents are selected, something we will discuss later. Next on the row are the coordinates and time of birth of the event the ray is coming out of followed by the angle with the $x$-axis. These are photons so they propagate at a speed $c=1$ in natural units, and at a $45º$ degree angle with the $z$-axis which is our time axis. This is always the case since we take the space components to be both the $\sin \phi$ and $\cos \phi$ with $\phi$ a randomly-generated angle in the 2d plane for the particle (we add $\pi$ to its corresponding anti-particle).\footnote{Again, remember this is what we do when we define our initial conditions/events.} Therefore:
\f
\theta=\arctan \left(\frac{\sqrt{\sin^2\phi+\cos^2\phi}}{1}\right)=\frac{\pi}{4}\,.
\ff

Next we have the past of the event defined by Eq.~\ref{2dpast}  and finally the family the event belongs to. These two tables are equal with regards to their structure and the information they provide. 

The ``allevent'' table, for all events, is simply the table that records the spacetime position of all the events organized by families so we can later plot them accordingly. We have a number of rows equal to the number of families, the first number in each row is the number of the family and after that the coordinates of created events are added to the row of their corresponding family.

\subsubsection{coordinates}

This is where we will use the parametric equations of the selected parent rays\footnote{We will sometimes refer to both rays and anti-rays as just rays or particles but remember that interactions are only between particles and anti-particles.} to find the spacetime location of their intersection. Each row represents a certain instant. We go forwards in time starting at the time of birth of the younger parent all the way to $T_{{\rm max}}$ plus this time in steps of the chosen $dt$. These two variables are actually very important and we will discuss them better when we discuss how we compute the intersections. So with each event we are creating this table with $T_{{\rm max}}/dt$ rows. We can easily see here the computational importance of these two parameters and how they will affect the speed of the program. Especially since they will also dictate, in the same way, the number of rows the ``distances" and ``Distances" tables will have. The computation of the interaction and therefore the new event is, as expected, what drains more computational power on the program and these are the two parameters that control it. 

They aren't the only ones responsible though. The radius we chose for the photons through $\bar{\sigma}$ also dictates how fast they will interact which in turn dictates how to choose $T_{{\rm max}}$ and $dt$. They can't have arbitrary values. If we have a really big $T_{{\rm max}}$ sandwich like structures will appear like we will see, since we allow too many windings to take place, meaning the photons are crossing the boundary many times before intersecting each other, this presents no problem though. Make it too small and there might not be any rays intersecting in this amount of time and this is a problem. 

With $dt$ we also have problems, if it is too small compared with the radius we will calculate the intersection with great precision but the program will take too long to run even for $1000$ events and we are interested in over $10^5$ events. Like we already discussed it can't be bigger than a diameter since we might miss an intersection and even being smaller than that it if we make it too big, we have poor precision in computing the coordinates for a new event.

The rows in this table consist of the particle and anti-particle's spatial coordinates $x$ and $y$, and then the time instant the coordinates correspond to. We use the ``Mod'' function to enforce our boundary conditions.

\subsubsection{``distances''}

In this code table named ``distances'' each row corresponds again to one instant of time and it is where we take the spatial distance in each axis between the rays. Care must be taken here. Since we have boundary conditions there are two possible distances in each axis between all rays at all times. The direct distance between them, let's call it $d_x$ for the $x$-axis for example, and the distance through the boundary that corresponds to, in this case, $L-d_x$ . We take the minimum of these two distances to be the actual distance. To understand this choice better let's think of what our boundary conditions do to our 2d square. When you apply boundary conditions on a line segment you turn it into a circumference, when you have a square you turn it into a torus. So picture two events anywhere on a torus, you can either take the longer path or the shortest path between them to be the distance. The most natural choice is to take the smallest path as the actual distance. We actually have to make this choice because if we don't the quasi-particles won't emerge since we won't be consistently choosing the closest pasts.

\subsubsection{``Distances''}

This code table named ``Distances'', again with each row corresponding to one instant of time, is just the Pythagoras theorem of the values of the previous table minus the $2\bar{\sigma}L$ ($\bar{\sigma}L$ is the actual radius) and the instant they correspond to. Since we are taking the value of two times the radius of the rays every time the values become negative we have an interaction and therefore a new event. So the first instant we find a negative value on this table we have an interaction, hence the importance of $dt$ in the precision of this computation. We removed the square roots in these calculations since they don't alter the result and this reduces computational time.

\subsection{Computing the intersections}

Now we'll be discussing how we find the intersections between our rays and therefore define new events. Computing and defining an intersection were the hardest parts of this work, they were the main problem to solve going from 1+1d to 2+1d. For now we will show how we compute the intersection and later we will talk about how we define them. We tried many things until we were able to get the results we wanted. We simplified where we could as much as possible since the goal was not to accurately define intersections. This is also, as to be expected, the most important and most demanding in terms of computational power part of the code. There are two ``While'' cycles, that we'll be calling the big ``While'' and the other just ``While'' which is inside the first. So inside the big ``While'' cycle we have our ``coordinates", ``distances" and ``Distances" tables. In the end we have the other ``While" whose purpose is to find the first negative negative value in the ``Distances" table. Remember that is when we have an interaction and therefore a new event.

We are envisioning our particles as circles or discs of radius $R$, $T$ on the code for thickness. This is necessary since we need to attribute some finite measure, cross-section, to our particles for interactions to occur. Not that 2d measure zero particles can't intersect on the plane but the probability of them doing so with random initial positions and directions is infinitesimally small. In other words, the subset of initial conditions that allows for the intersection of two 1d measure lines in the plane has measure zero. Therefore the probability of picking such initial conditions is zero. We have particles and anti-particles, one pair leaving each event. Again, we will only allow intersections between particles and anti-particles. The initial events are created with random positions in our space at global time $t_{\rm Minkowski}=0$. We generate the particles direction in the 2d plane $\phi$ randomly and their corresponding anti-particles to be traveling in the opposite direction $\pi + \phi$.

Due to our boundary conditions the best and simplest way to find intersections is through the parametric equations of the particles. But this raises a problem since our particles are not points. The natural candidate to represent our particles as a point is the centre of each circle. So we use the parametric equations to propagate this point taking the positions of the corresponding events to be the initial positions of these points and interactions will be defined when the spatial distance between two lines is smaller than a diameter, meaning as soon as they touch there is an interaction, ideally when the two circles are tangent to each other. 

So the first step is to create a table were we compute the position of each ray at time $t$. That's the ``coordinates" table. When computing the coordinates it is important to remember that the parametric equations of each ray must be time shifted to the time of birth of the corresponding event they belong to, otherwise it would be like evolving all rays from the initial time. Another thing to keep in mind is that we obviously don't want a baby event to be born before its parents, that is not discausality, that's nonsense. We easily avoid this by having the time we start computing the coordinates starting only at the time of birth of the youngest parent, which is the maximum between the times of birth\footnote{$ct$, for collision time, is the name of the variable in the code for the time at each event is born} of both parents. We will then compute the coordinates for a certain amount of time $T_{{\rm max}}$. This also ensures that time keeps moving forward, otherwise we would be stuck in the $0$ to $T_{{\rm max}}$ interval. After that we just compute the distance in each axis between the rays using the ``distances" table the way we described. Then use the Pythagoras theorem in the ``Distances" table to get the actual spatial distance and we subtract a diameter from it. So now all we need is to look in this table for the first negative value, and that is the purpose of the final ``While''.

There are a few difficulties to discuss. First we don't know when the particles will intersect each other, so for how long should we compute the coordinates or, in other words, how big should our $T_{{\rm max}}$ be? What if the $T_{{\rm max}}$ we selected isn't enough, what should we do? Should we allow $T_{{\rm max}}$ to be increased if needed? Should we try and find another pair that intersects in the selected $T_{{\rm max}}$? We think we can all agree that trying to find or even guess, based on the initial conditions what would be a $T_{{\rm max}}$ that would be enough for computing as many intersections as we wanted is impossible. Even if we could, it would be computationally very heavy since it would be the same value for all interactions and for most of them wouldn't take that long, like in cases when the rays are almost colliding when we start computing the coordinates. So we either allow, when necessary $T_{{\rm max}}$ to update itself or pick another pair to interact whenever we run out of time. That is why we need all the algorithm to find an intersection inside a ``While'', the big one. We tried both of these hypothesis, and computational power ruled out $T_{{\rm max}}$ updating itself with each iteration of the big ``While'' until the intersection was found, so we select a $T_{{\rm max}}$ and if it isn't enough we pick different parents until we find a suitable pair for the chosen $T_{{\rm max}}$. The way we re-selected parents is simple but program specific so we will postpone this to the next subsections.

So how do we start the big ``While'' cycle and why? We begin by defining two variables, the time of birth of the youngest parent and the collision time $ct$ of the event to be created. We initially set these two variables equal so the cycle can start. So we say, while $ct$ is smaller or equal than the time of birth of the younger parent we keep choosing parents until we find a pair that intersects during our $T_{{\rm max}}$. This has to be done because of a glitch that was constantly appearing. 

Let us give an example of what could go wrong when the first baby event is created. This will not happen just to the first event being created, but this makes it simpler to explain. So let's say we selected both of our parent events, \textbf{A} and \textbf{B}. The particle from event \textbf{A} and the anti-particle from event \textbf{B}, could be the other way around. And let us say that at time $t=0$ which is, in this scenario, is the time at which the coordinates will start to be computed since it is the time of birth of the youngest parent, the two rays are already overlapping. That will mean that the first element in the ``Distances" table will be negative which means $ct$ will be selected to be zero. Now, as we said, while $ct$ is smaller or equal than, in this case $0$, we keep choosing parents until we find some that interact in the chosen $T_{{\rm max}}$. But let's say we didn't do that. Well, an event would be created at $ct=0$, event \textbf{C} with the particle from event \textbf{A} and the anti particle from event \textbf{B} coming out of it. Now, this third event is already overlapped with both events \textbf{A} and \textbf{B} and these events still have respectively an anti ray and a ray, so they will still be able to interact both between themselves and with event \textbf{C} creating two more events which will interact between each other all at the same time $t=0$ and at the same point in space. This process will continue to the end of the run with all created events occupying the same spacetime point. 
This a real problem when we are using a law that chooses the parents based of the minimum difference between pasts and is obviously a glitch which we solved by defining $ct$ to be equal to the time of birth of the youngest parent in the beginning and only stop when $ct$ is bigger from it. 

Finally, when looking for the first negative value in the ``Distances" table, we need to make sure the program actually returns the first negative value, so we break the ``While'' cycle used to go through the ``Distances'' table after it finds a negative value. After the big ``While'' we define a variable called ``maxt" with the purpose of recording the highest $ct$ of the run for plotting purposes. Then all that remains is for us to update our thick present and ``allevent'' table.

\subsection{Adding and removing events}

 The first thing we do is update the ``allevent'' table. To know the family of the parent particle (remember its always the particle that gets to be the final parent) we just access the information of the ``livev1'' table, it will be the last element in the row. Having the right family for the new event we simply add it to the end of its corresponding row. Next we move our attention to the ``livev'' tables adding the new events to them in the last position. Finally we delete the used rays from both ``livev'' tables.
 
 We are now in a better position to understand why we divide the thick present into two tables, which we justified by having particles and anti-particles, and how choosing only the particles or the anti-particles family to inherit a new event ensures conservation of families throughout the run. Saying that the final parent is always either the particle or the anti-particle ensures the number of different families in the corresponding table remains conserved since we add and delete an event from the same family in each iteration. What happens on the other table is of no importance since it is the first that determines the final family. Initially we only had one table of photons, still with two leaving each event but both in random directions (not opposite to each other). There is no way of enforce the conservation of families like this.

 For instance, in the 1d case it was always the right-moving photon that determined the final family and this ensured the number of families available for interaction remained conserved. What we are doing is replicating this mechanism in the 2+1d case. Eventually we will want to see some families stop interacting or interacting much less, but not because they are no longer present in the thick present and therefore no longer available for interaction. If we don't enforce conservation of the number of families, as a family gets more and more events in thick present compared to the others it gets a higher probability of being selected to host new events in the future. As time progresses only one family will end up in the present and therefore only that family will be able to interact with itself. We will explain better in the next chapters how this happens specifically in each program, but this is the main cause.

\subsection{Randomly choosing our parents}

This algorithm is to test the robustness of the code, meaning that we use it to test if in each run we end up with an almost even distribution of events between the families when the parents are selected completely randomly and the pasts play no role. That will mean we obtained conservation of families in the thick present. In order to do this, before the big ``While'' cycle we define two tables which we called ``\textbf{a}" and ``\textbf{b}" for lack of a better name. These tables just index the rows of the ``livev'' tables. 

Then we create a table named ``parents" that takes all the possible combinations of pairs of indices we can have. So the first two tables have simple purpose, to index the rows of ``livev1'' and ``livev2'' respectively and the ``parents" table creates all possible pairs of indices so we can then use the ``RandomChoice'' function to randomly pick a pair from the ``parents" table. We need to take care when doing this, because we don't want rays from the same event interacting so, to make sure, we use an ``If" to remove the the pair from the ``parents" table in case they are from the same event and re-select randomly another pair. At the end of each iteration of the big ``While" cycle if the pair selected didn't intersect we also remove this pair from the ``parents" table.

We work with the tables ``\textbf{a}", ``\textbf{b}" and ``parents" precisely to have this freedom to remove at will events whenever needed without affecting further iterations. 

We want to remove the pairs to make sure that if we run out of parents, meaning that there are no available rays able to interact before our $T_{{\rm max}}$, our big ``While'' cycle doesn't keep on running forever. So at the end of each computation we check to see if our $ct$ is still equal to the time of birth of the younger parent, given that this would mean no collision was found. If that is the case, we remove the used pair from the parents table, since that pair is, in effect, useless for this iteration. Then, at the beginning of our big ``While'' cycle we check if the length of the ``parents" table is zero and if it is we abort the run. 

We will now take a closer look at how the thick present can get dominated by only one family if we don't select the final parent the way we do. Let's say we are selecting the final parent randomly out of the two parents which, in effect, we did and is, in practice, equivalent to everything else we tried to solve this issue. Let's say we have $N$ families and only one table for the thick present. In this case, each row will have the two rays. When we create the first event, everything being random any family is equally likely to inherit the first new born. But then, for the second creation, and since only two rays from two different events have been used there will be eleven events in the thick present with two events belonging to the same family\footnote{The two parent events still have one particle each available for interaction.}. Now this family has a higher probability of being selected to interact the next time and therefore hosting even more events. This process will happen until one family has won and all events in the thick present belong to this family. Nothing determines the winning family: it is only a matter of chance. So again having two tables for the thick present and ensuring that only the family of only one of the tables is selected to host a new event (and making this choice consistently) makes it so that in that table, which is the table that matters for our purpose, the number of families there remains constant throughout the runs since within each iteration/creation of a new event, we remove and add a ray coming from the same family.

\subsection{Deterministic version}

This program is to study better the emergence of quasi-particles, which happens more quickly without randomness. This is because the system gets caught by the limit cycles quicker since it only needs to overcome the random input of the initial conditions. In this fully deterministic version we use our rule, which chooses the parents by extremising the differences between their pasts using 
\f
D_{IK}^{(2d)}=|past_I^2-past_K^2|\,.
\label{pastdiff2d}
\ff
For that we create a table, ``DIK'' that takes the absolute differences between the pasts just like in the 1d case Eq~\ref{pastdiff1d}. In each row we have that difference and, for computational purposes the events that correspond to that difference. We just need to make sure we are not taking the differences between equal events and we will be sure two rays coming from the same event won't intersect. We do that simply by using the information from both ``livev'' tables to make sure the two rays we are computing the difference for don't come from the same event. We then select the parents for whose difference in pasts is the smallest, again just like in the 1d case.

 If the two selected parents don't collide in the time we want, we select the parents whose difference is the second minimum value and so on. We accomplish this by using the function ``RankedMin". We also want to make sure, just as in the previous case, that if we run out of rays the big ``While'' cycle doesn't run forever. In this case it is much simpler. We simply define a variable we called, again with lack of a better name \textbf{``z''}, that will count the number of iterations of the big ``While'' and when it's value reaches the ``Length" of the ``DIK'' table it means we tried all possible combinations and it is therefore time to abort. We also use \textbf{``z''} to define the ``RankedMin'' function, first minimum, second minimum etc.

The domination of the thick present by only one family in this case is a bit different,  since the parents aren't being selected randomly. It is the same reasoning as before, one family starts to dominate eventually and after some time only that family exists in the thick present. What makes that happen here isn't just probability but also time. Since we are taking the pasts in our restricted space, due to the boundary conditions, time keeps moving forward but the spatial coordinates are restricted to the dimensions of our space. So after a few runs the pasts get dominated by the time coordinate and the families that have already interacted have a bigger probability of being picked for future interactions since these are the ones for which time has evolved more and so the difference between them will be the smallest. This phenomenon is part of the dynamics and this is what is expected to happen. However the fact that one or more families have more events in the thick present relative to the others because the number of families isn't being preserved, not only will speed up the process but also can make it so that a family that could end up emerging gets eliminated from the present without having that chance. This would naturally alter results of the runs because it is changing the dynamics in the set, so it is very important to preserve the number of different families in the present through out the run.

\subsection{Mixing randomness and determinism} 

This version is the final product where we introduce a variable called ``random" that controls the probability of the parents being selected randomly. This program is just a mix in one big ``If" that puts together the two big ``While'' cycles of the other two programs. So we choose ``random" to be a number between 0 and 1 and if a created pseudo-random is smaller than ``random" the parents are selected randomly. So in effect, the number we choose for ``random" is the probability of selecting the parents randomly. 

There are two ways of imposing time irreversibility, through the surjectivity of the laws, or by allowing probability to enter the picture. We choose randomness to introduce time irreversibility into the model. Everything else about this program is the same as the other two. We expect the other two programs to be limits of this \textbf{final program} when ``random" approaches, respectively, $1$ and $0$ and we will see if that is the case in the next chapters.

This version of the code is presented in the Appendix.

\section{Robustness of the Code and Parameter Control}

In this section we present the results of the random, deterministic and mixed case. We present several runs to show the effect of the parameters on the run and discuss it. This serves to show our control over the code. We will also confirm if the deterministic and random cases are limits of the mixed case. It is expected that, in the limit of very small randomness, the results of our final program approach those of the fully deterministic version. On the other hand, as randomness tends to 1, we would expect the results to approach those of the fully random version. We would like to emphasize that in this part we merely present our results and discuss the effect of the parameters leaving the discussion of the results for later. 

\subsection{The deterministic limit}

\begin{figure}[t]	
	\centering
	\includegraphics[width=150mm]{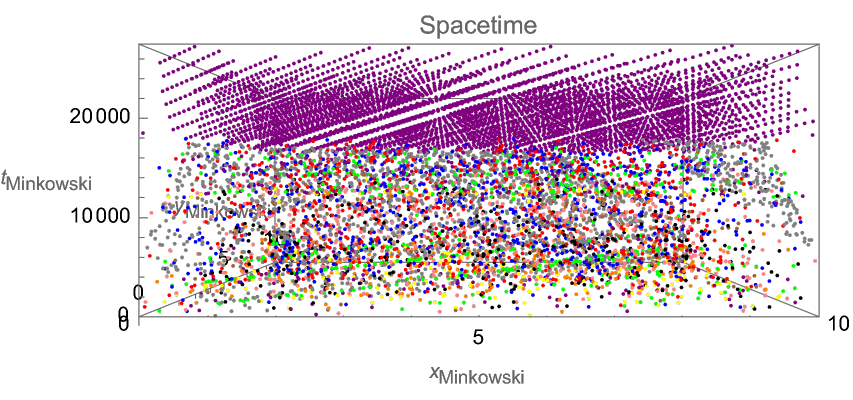}
	\caption{A run from the fully deterministic program, where we select the parent event according to our time evolution rule that picks the events with closest pasts. Parameters: $\bar{\sigma}=0.05$, $dt=1\times \bar{\sigma}$, $T_{{\rm max}}=1 \times L$, $L=W=10$.  10 Families , 10 000 Events. Each color corresponds to one family. One can clearly see on the plot a phase transition from a disordered phase where all families are interacting to an ordered phase where only one family interacts. This characterizes the emergence of quasi-particles and therefore the emerge of the time-symmetric phase of the run.}
	\label{fullydet}
\end{figure}

Figure~\ref{fullydet} shows a run with 10 families and 10 000 events of the deterministic code, with each color representing one family. We can observe a clear phase-transition from a random, disordered phase to an ordered crystal like structure phase. This crystal structure characterizes the emergence of quasi-particles in 2+1d. The main objective in this thesis was obtaining this phase transition and control it with the parameters. We don't want a crystal without a random phase nor do we want only a random phase. Just like in the 1d case, the disordered phase will represent the time irreversible part of the run with approximate time reversible dynamics emerging after a while, hence we want both phases to be clearly present. So let's see how the parameters affect the runs first, starting with the number of families.

Even though this is the fully deterministic case there is a random input that is always present. The input comes from the randomly selected initial conditions, the initial positions and directions. So the bigger the number of families the bigger the random initial input in the system. Therefore the longer it will take for the crystal to form. This is because the crystal phase characterizes time symmetric evolution and the disordered phase the asymmetric time evolution generated by the random input. So before the crystal emerges the system needs to overcome all the randomness, in this case created at the beginning of the run. The bigger this random input the longer the system will take to settle in a limit cycle characterized by the crystal ordered phase.

Before moving any further we would like to remark that, like with any chaotic system, these runs are very sensitive to the initial conditions. Therefore, the behaviours we are discussing in this chapter are to be taken as on average behaviours. Runs with same parameters can differ a lot, but naturally, there is a pattern. 

\begin{figure}[htpb]	
	\centering
$\begin{array}{cc}		
		\includegraphics[width=80mm, height=55mm]{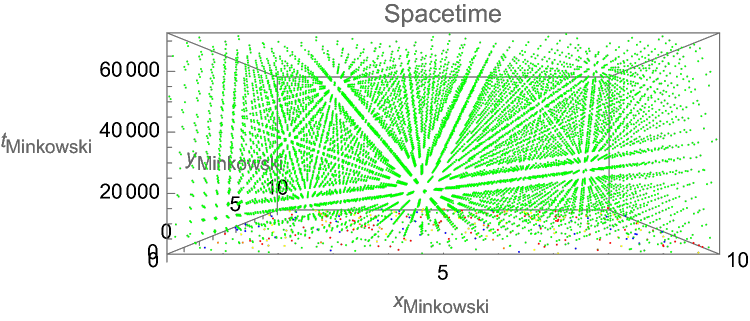} &
		\includegraphics[width=80mm, height=55mm]{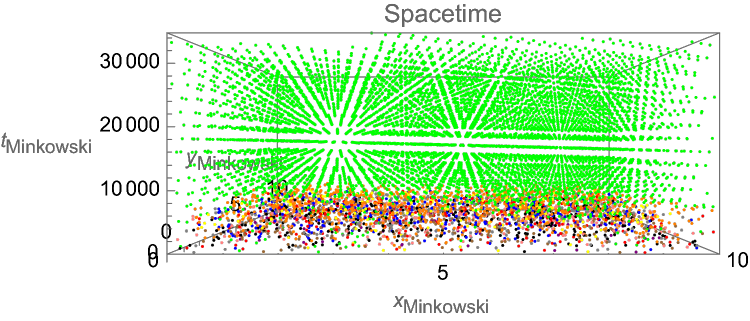}\\
		\textbf{(a)} & \textbf{(b)} \\
		~\\
		\includegraphics[width=80mm, height=55mm]{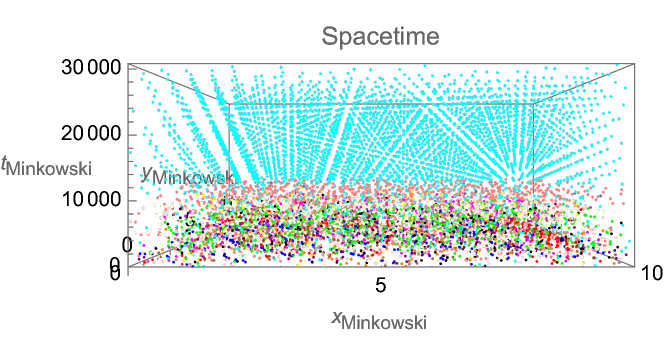}&
		\includegraphics[width=80mm, height=55mm]{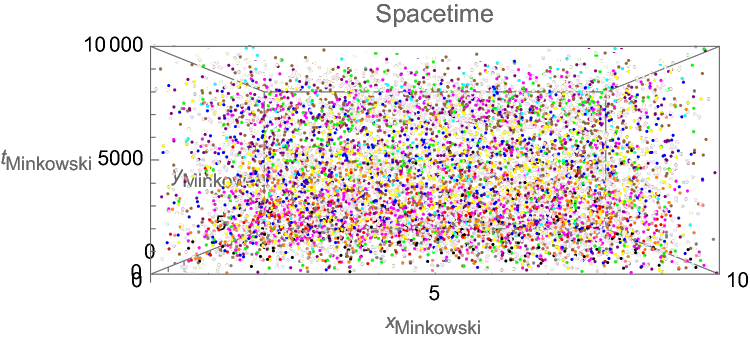}\\
		\textbf{(c)} & \textbf{(d)} \\
		~\\
		\end{array}$
	\caption{A run from the fully deterministic families to compare the effect the number of families has on the run. We only vary the number of families. Parameters: $\bar{\sigma}=0.1$, $dt=1\times \bar{\sigma}$, $T_{{\rm max}}=1 \times L$, $L=W=10$. 
	All the plots have the same number of events, 10 000. Plot \textbf{(a)} has 5 families, Plot \textbf{(b)} 10 families, Plot \textbf{(c)} 15 families, and Plot \textbf{(d)} 20 families. We are clearly able to see an increase in the time it takes for the crystal to emerge, and in the last plot the run was not long enough for the crystal to form. This is because a bigger number of families equates to bigger random input at the start of the run.}
	\label{fullydetcomparefam}
\end{figure}

\begin{figure}[t!]	
	\centering
	\includegraphics[width=150mm]{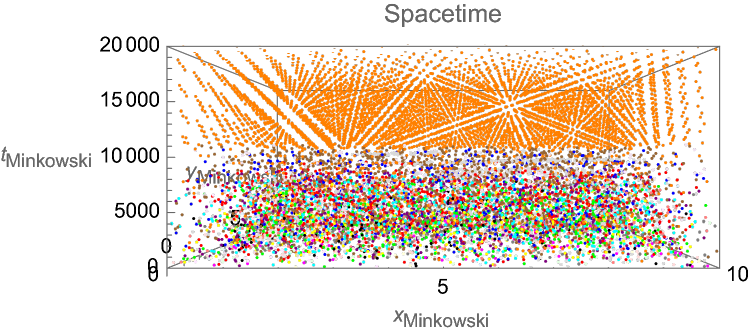}
	\caption{A run of the deterministic program with 20 families. Here we want to show that even if we have a large number of families, given enough time the system will settle and form the crystal. Parameters: $\bar{\sigma}=0.1$, $dt=1\times \bar{\sigma}$, $T_{{\rm max}}=1 \times L$, $L=W=10$, 20 families, 100 000 events. We only present the run up to 20 000 units of time so the phase transition is clearer.}
	\label{fullydetlargefam}
\end{figure}

Figure~\ref{fullydetcomparefam} shows four runs with 5, 10, 15 and 20 families. We can clearly see that the crystal takes longer and longer to emerge and with 20 families the run wasn't even long enough for it to form. Eventually the crystal is always formed in the deterministic version and we can see that in Figure~\ref{fullydetlargefam}.

\begin{figure}[t]
	\centering
$\begin{array}{cc}		
	\includegraphics[width=80mm, height=55mm]{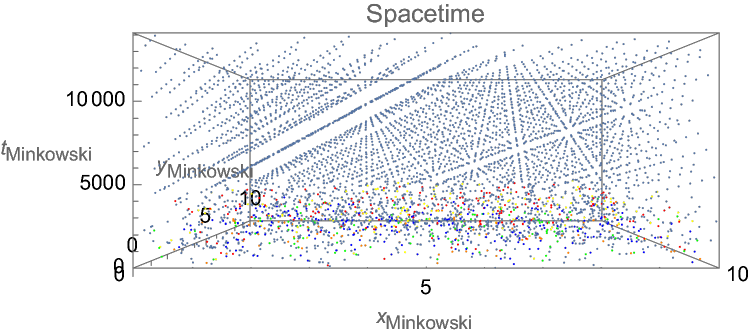} &
	\includegraphics[width=80mm, height=55mm]{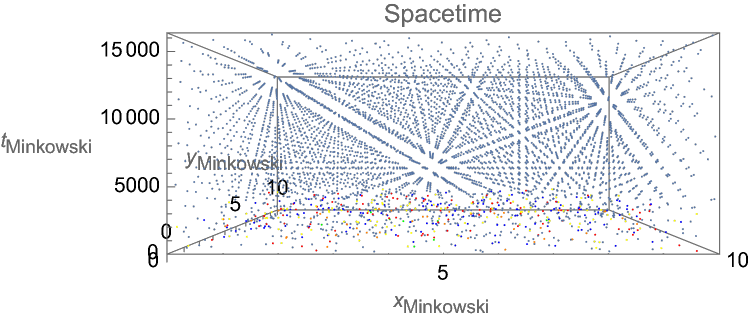}\\
		\textbf{(a)} & \textbf{(b)} \\
		~\\
	\includegraphics[width=80mm, height=55mm]{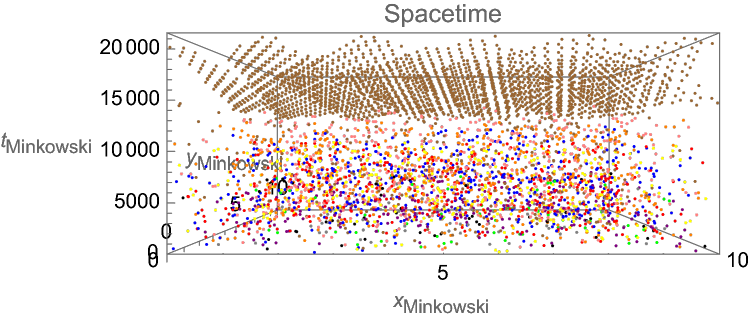} &
	\includegraphics[width=80mm, height=55mm]{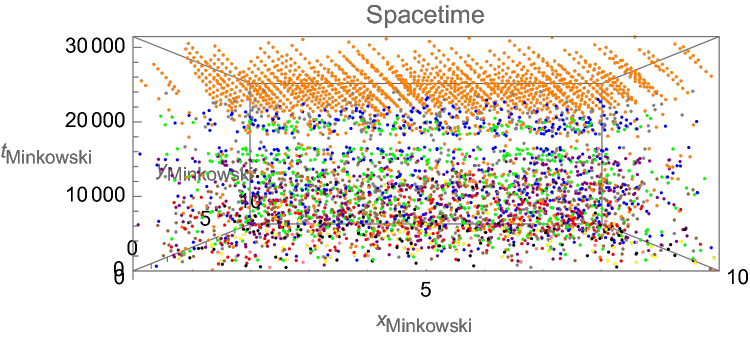}\\
		\textbf{(c)} & \textbf{(d)} \\
		~\\
		\end{array}$
	\caption{A run from the deterministic program to compare the effect of the variable $T_{{\rm max}}$ on the runs. We only vary $T_{{\rm max}}$. Parameters: $\bar{\sigma}=0.1$, $dt=1\times \bar{\sigma}$, $L=W=10$, 10 families, 5 000 events. Plot \textbf{(a)} shows a run with $T_{{\rm max}}=1\times L$, Plot \textbf{(b)}  $T_{{\rm max}}=5\times L$, Plot \textbf{(c)} $T_{{\rm max}}=20\times L$,  and Plot \textbf{(d)} $T_{{\rm max}}=20\times L$. It can clearly be seen that as the parameter $T_{{\rm max}}$ is increased the runs last for longer Minkowski time and the emergence of the crystal happens later.}
	\label{comparetmax} 
\end{figure}

Let's now take a look at the effect $T_{{\rm max}}$ has on the emergence of quasi-particles. First of all, as $T_{{\rm max}}$ gets bigger particles have more time to interact. Therefore time will go further due to particles going through more windings before interacting. This will, in principle, make the emergence of quasi-particles appear later. Figure~\ref{comparetmax} shows runs with 10 families and 5 000 events where we only vary $T_{{\rm max}}$, that takes the values 10, 50, 100 and 200. One can clearly see in the plots that the bigger the $T_{{\rm max}}$, the longer the runs and the longer it take for the crystal to emerge. One can also note the sandwich effect we talked about in plot \textbf{(d)} very clearly around time 20 000. This effect, as we already said, increases with an increasing $T_{{\rm max}}$ and causes no problem to the dynamics.

\begin{figure}[htpb!]
	\centering
$\begin{array}{cc}		
	\includegraphics[width=80mm, height=55mm]{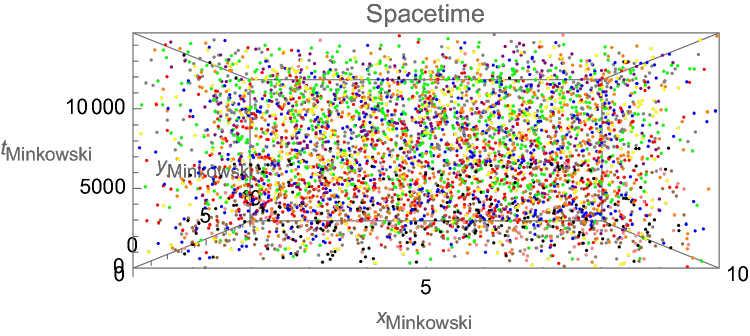}&
	\includegraphics[width=80mm, height=55mm]{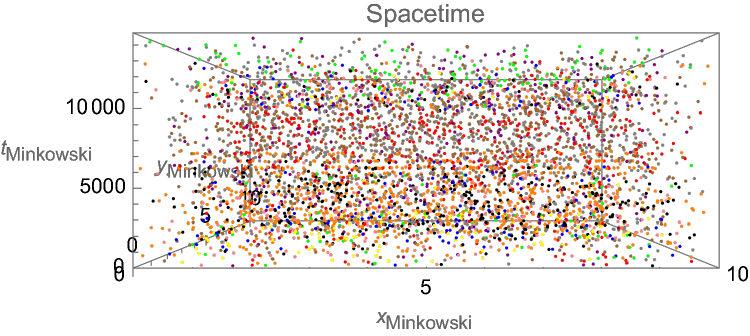}\\
		\textbf{(a)} & \textbf{(b)} \\
		~\\
	\includegraphics[width=80mm, height=55mm]{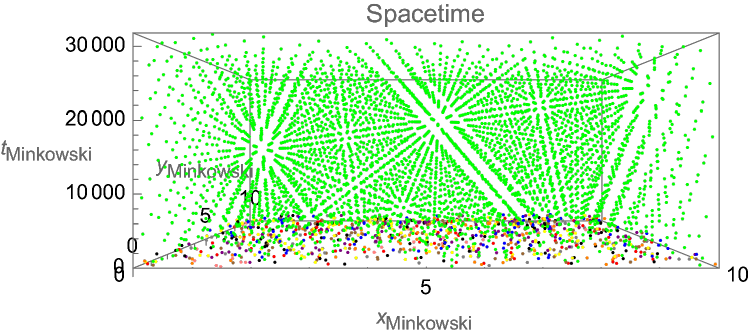}&
	\includegraphics[width=80mm, height=55mm]{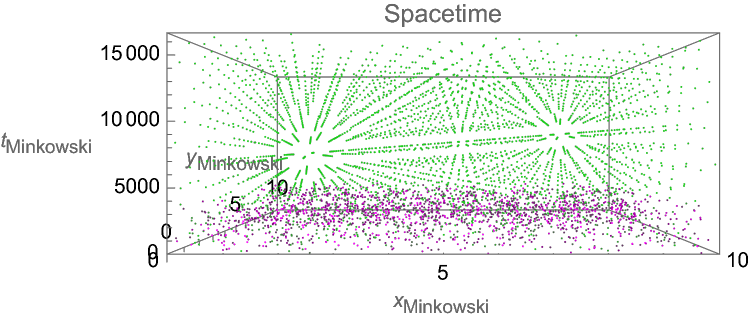}\\
		\textbf{(c)} & \textbf{(d)} \\
		~\\
	\includegraphics[width=80mm, height=55mm]{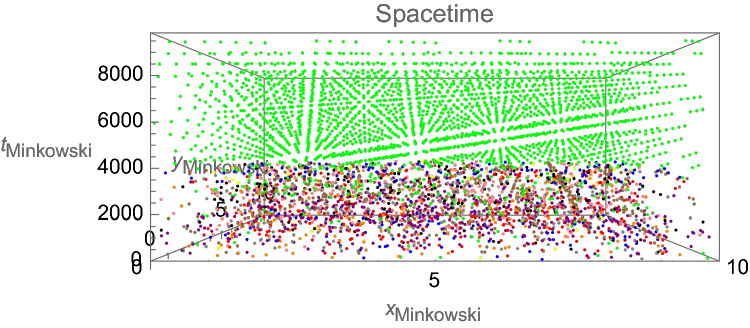}&
	\includegraphics[width=80mm, height=55mm]{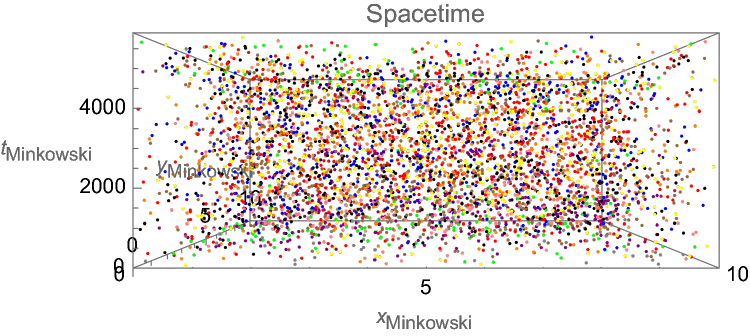}\\
		\textbf{(e)} & \textbf{(f)} \\
		~\\
		\end{array}$
	\caption{Run of the deterministic program to see the effect of $\bar{\sigma}$ on the runs. We only vary $\bar{\sigma}$. Parameters: $dt=1\times\bar{\sigma}$, $T_{{\rm max}}=1\times L$, $L=W=10$. Plots \textbf{(a)} to \textbf{(f)} show $\bar{\sigma}=0.05$, $0.075$, $0.1$, $0.15$, $0.2$, and $0.25$ respectively. As we increase the radius the crystal forms more quickly, since given that the particles are bigger they will meet faster. However there is a breaking point due to the way interactions are computed. This presents no problem since it only occurs when particles become too big, in which case the run loses its interest.}
	\label{crosssectioneffect}
\end{figure}

We've already seen what happens when we change the number of families and $T_{{\rm max}}$, so now we will see what happens when we change the cross-section, $\bar{\sigma}$. We won't change the time step because, as we discussed it is good to keep it equal to $\bar{\sigma}$. 

Reducing the thickness will increase, in general, the amount of time it takes for particles to interact and if we don't have a big enough $T_{{\rm max}}$ or a big number of families eventually no pair will be found to interact and the run will be aborted, so we need to look out for that. Increasing the radius will reduce, in general, the amount of time taken for particles to interact, but if we go too far, it raises problems as we shall see. Figure~\ref{crosssectioneffect} clearly shows this.

\subsection{The random limit}

The main thing here is to check that when the parents are random we obtain an almost even distribution of events among the families. And we want this to see if the number of different families in the thick present gets preserved. 

When we estimate the standard deviation from the runs the result will be subject to sampling noise. For the reasons mentioned below we interpret a result within a factor of 2 of the theoretical value as consistent with a random (Poisson) distribution.

The Poisson distribution is the adequate to describe how many events go into each family. We imagine we throw one object into a randomly chosen box, which represents a family. After some amount of time we will want to know how many events are in each of the boxes. In each individual box there is, at any given time 1 in $N_{pasts}$ (number of pasts/families) chance that the event will end up in the box. There is then an expected number of particles in that box: $1/N_{past}$ of the total number of events, as well as some distribution around that number, caused by randomness. This forms the classical counting problem for which the poison distribution is the appropriate description.

The observed distribution of events is more broadly distributed than we are predicting in a Poisson distribution. When $T_{{\rm max}}$ is increased the ratio approaches the theoretical limit. There remains some correlation from event to event leading to a somewhat wider distribution than a pure random process, for the length of runs carried out. Roughly speaking we have considered that the interval 0.5-2 denotes. If the process was genuinely random we might expect to get a actual distribution that was within a factor of two of the theoretical value. The fact is we don't have enough families to estimate the standard deviation very well, our set of pasts consists typically of 5-10 families.

\begin{figure}[t]
	\centering
	\includegraphics[width=150mm]{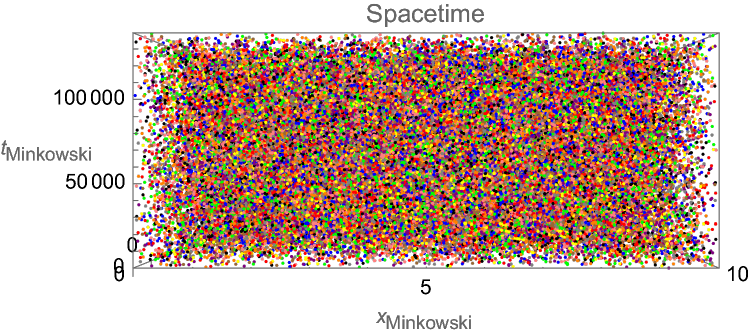}
	\caption{A run from the fully-random program. Parameters: $\bar{\sigma}=0.1$,  $dt=1\times \bar{\sigma}$, $T_{{\rm max}}=1\times L$, $L=W=10$, 10 families, 100 000 events. No structure arises and we get an approximately even distribution of events between the families.}
	\label{fullyrandom}
\end{figure}

Figure~\ref{fullyrandom} shows a run where the parents are selected completely randomly. We got the following distribution of events among the families: 

\begin{multicols}{2}
	\begin{itemize}
	\centering
	\item Family 1: 10.7\%
	\item Family 2: 10.8\%
	\item Family 3: 11.9\%
	\item Family 4: 10.2\%
	\item Family 5: 9.8\%
	\columnbreak
	\item Family 6: 10.0\%
	\item Family 7: 7.7\%
	\item Family 8: 10.4\%
	\item Family 9: 11.8\%
	\item Family 10: 6.7\%
	\end{itemize}
\end{multicols}

The results are convincing; we can say that the number of families gets preserved since throughout the run no family stopped interacting. However one might say that even though this is an approximately random distribution, for so many events one would expect the distribution to be even smoother. One could then say that maybe family 10 for instance having a much lower percentage than the others could have stopped interacting at some point. The problem here is  that $T_{{\rm max}}$ is too small and this is actually the only important thing about parameter control here. Let's try to increase it to see if we get better results.

\begin{figure}[t]
	\centering
	\includegraphics[width=150mm]{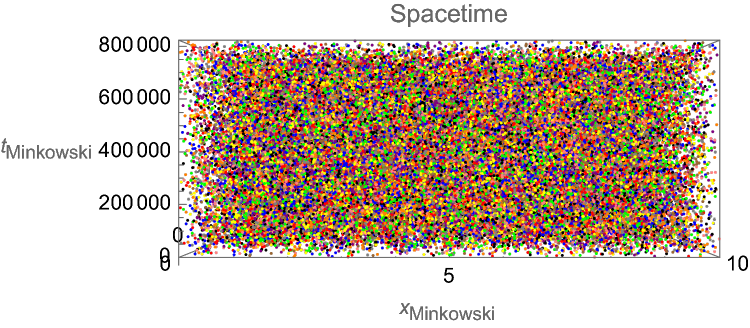}
	\caption{A run from the fully-random program to show the effect on the distribution of events among families given by an increase in $T_{{\rm max}}$. Parameters: $\bar{\sigma}=0.1$, $dt=1\times \bar{\sigma}$, $T_{{\rm max}}=10\times L$, $L=W=10$, 10 families, 100 000 events. Again no structure forms and we get a more even distribution.}
	\label{fullyrandomdist} 
\end{figure}

\begin{multicols}{2}
	\begin{itemize}
		\centering
		\item Family 1: 10.1\%
		\item Family 2: 9.9\%
		\item Family 3: 10.5\%
		\item Family 4: 9.5\%
		\item Family 5: 9.6\%
		\columnbreak
		\item Family 6: 9.3\%
		\item Family 7: 10.8\%
		\item Family 8: 10.0\%
		\item Family 9: 9.7\%
		\item Family 10: 10.1\%
	\end{itemize}
\end{multicols}

Clearly the results\footnote{The percentages of figure~\ref{fullyrandomdist} don't add up to exactly 100 because the values were rounded to one decimal place like in the previous case.} of figure~\ref{fullyrandomdist} are much better. The fact that we have to keep changing our parents a lot makes it so that some rays and anti-rays become more probable to be parents. Due to the initial conditions, there will always be pairs of rays and anti-rays that collide faster than others and these will be favoured if we are not willing to increase $T_{{\rm max}}$. This reinforces the idea that, in the deterministic case, the dominance of a family after the crystal is formed is due to dynamics and initial conditions. We will see in the discussion of the results of the deterministic case that sometimes more than one family dominates at the same time which again tell us that the program isn't forcing one family to interact more than the others, in other words this is not a coding issue but a dynamical consequence. 

\subsection{Introducing randomness to determinism}

\begin{figure}[t]
	\centering
	\includegraphics[width=150mm]{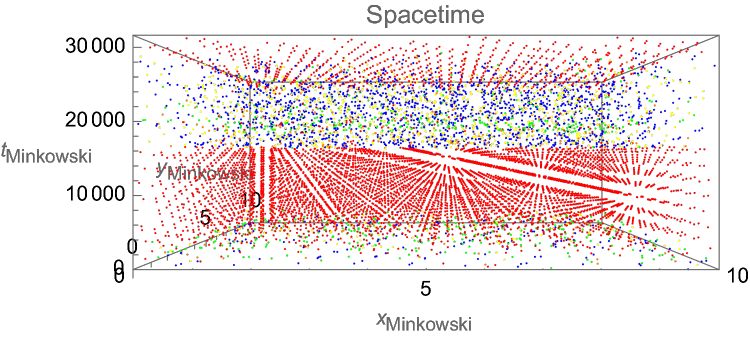}
	\caption{A run of the final program. Parameters: $\bar{\sigma}=0.1$, $dt=1\times\bar{\sigma}$, $T_{{\rm max}}=1\times L$, $L=W=10$, chosen randomness: 0.1\%, 14 randomly generated events, actual randomness 0.14\%, 5 families, 10 000 events. We can see the crystal emerging and getting fully formed just like in the deterministic case. Unlike the deterministic case however it gets broken.}
	\label{runfinal14} 
\end{figure}

We shall now show what happens when we introduce randomness to our deterministic dynamics. Figure~\ref{runfinal14} shows a plot with 5 families and 10 000 events. We can again see a phase transition but now the crystal gets broken after some time!

\begin{figure}[htpb]
	\centering
	$\begin{array}{cc}
	\includegraphics[width=80mm, height=55mm]{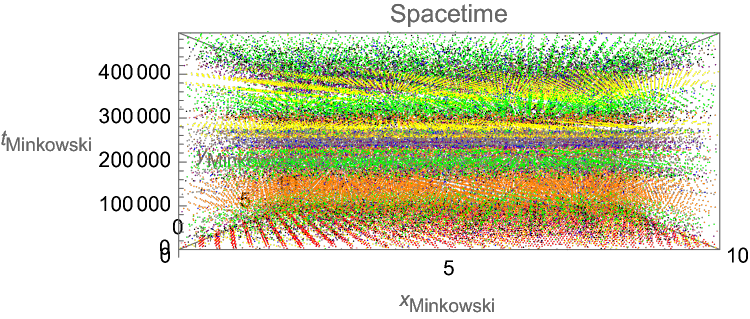}&
	\includegraphics[width=80mm, height=55mm]{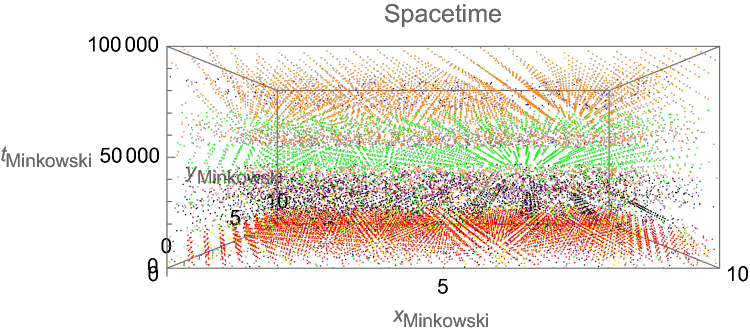}\\
		\textbf{(a)} & \textbf{(b)} \\
		~\\
	\includegraphics[width=80mm, height=55mm]{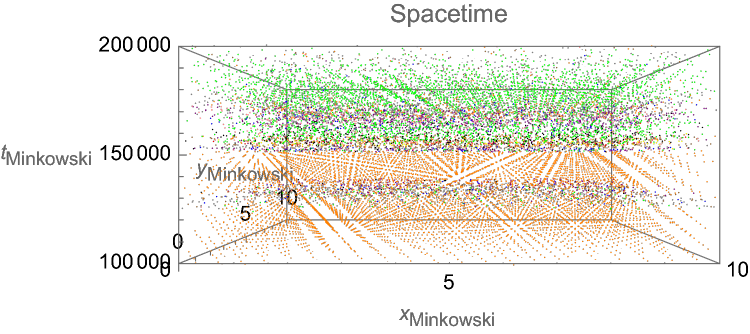}&
	\includegraphics[width=80mm, height=55mm]{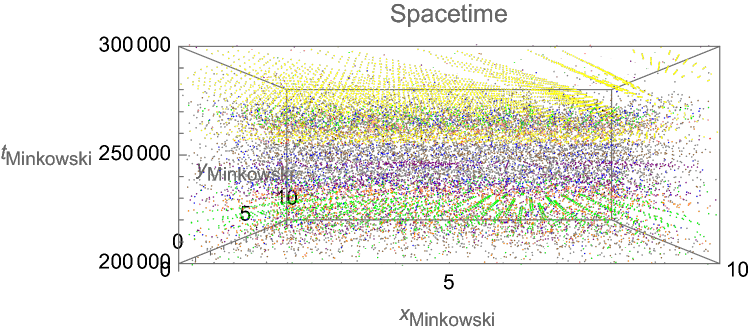}\\
		\textbf{(c)} & \textbf{(d)} \\
		~\\
	\includegraphics[width=80mm, height=55mm]{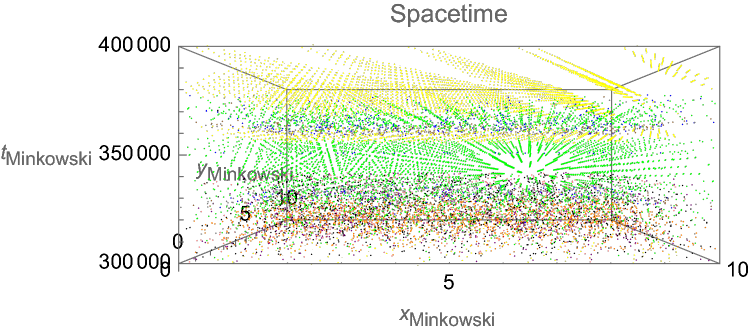}&
	\includegraphics[width=80mm, height=55mm]{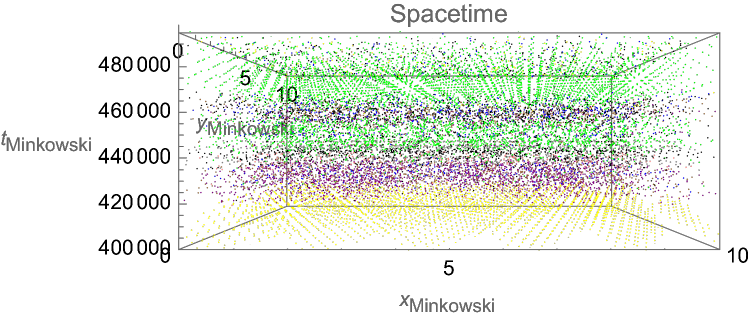}\\
		\textbf{(e)} & \textbf{(f)} \\
		~\\
		\end{array}$
	\caption{A run of the final program with more events to show the formation and breaking of crystals of different families. Parameters: $\bar{\sigma}=0.1$, $dt=1\times\bar{\sigma}$, $T_{{\rm max}}=1\times L$, $L=W=10$, chosen randomness: 0.1\%, number of randomly generated events 78, actual randomness=0.078\%, 10 families, 100 000 events. Plot \textbf{(a)} shows the complete run, with the following panels showing chunks of 100 000 units of Minkowski time each so the structures can be seen readily. We can see that different families form the crystal. This is again a great proof that there is conservation of families in the thick present.}
	\label{runfinalmore} 
\end{figure}

 The crystal forms and then we go back into a disordered phase again and then back again to a crystal phase. We would like to immediately point out two things. The first thing we can say after a lot of runs is that the system is highly sensitive to probability, which is to be expected since in this version the amount of random input put into the dynamics is part of the initial conditions. For instance 1\% is to much in general to get a fully formed crystal so one might think 0.1\% is too small but in fact, it isn't. The second is that there is no need for the same family to reform the crystal. In figure~\ref{runfinal14} the red family formed a crystal initially and later it was the red family trying to forming the crystal again. This does not need to happen and if it did happen it meant families weren't being conserved. 
 
Figure~\ref{runfinalmore} shows a run of the final code with 0.1\% of randomness but were more events were produced. There are clear formations of crystals from different families as time progresses. We can see clearly the competition ceasing, and one or more families trying to form crystals, but then, eventually they get broken. Then follows a disordered phase until the next crystal forms. When a crystal is broken its like a reset of the system and it takes some time before it settle again in a limit cycle. This is nothing like the fully deterministic case where when crystals were fully formed, they never got broken.

\begin{figure}[htpb]
	\centering
	$\begin{array}{cc}
	\includegraphics[width=80mm, height=55mm]{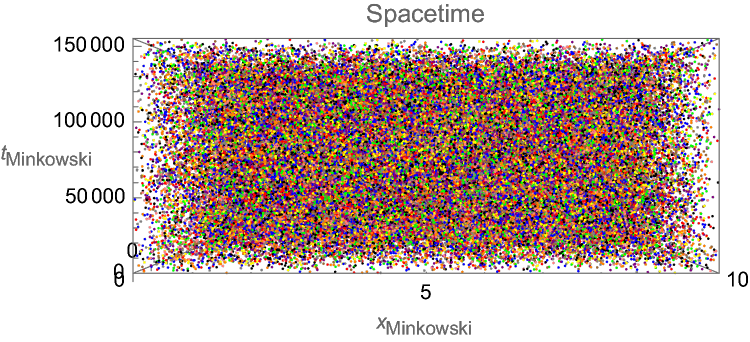}&
	\includegraphics[width=80mm, height=55mm]{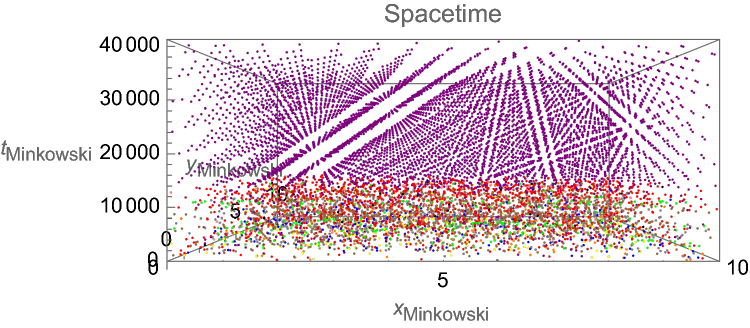}\\
		\textbf{(a)} & \textbf{(b)} \\
		~\\
	\includegraphics[width=80mm, height=55mm]{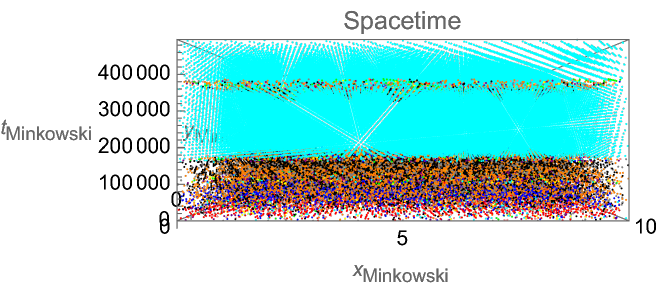}&
	\includegraphics[width=80mm, height=55mm]{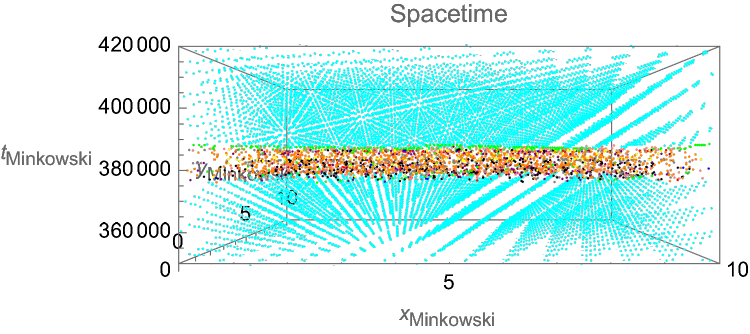}\\
		\textbf{(c)} & \textbf{(d)} \\
		~\\
		\end{array}$
	\caption{Three runs of the final program to see if we obtain the results of the previous programs to check if, in fact they are limits of the first.\newline
		 Plot \textbf{(a)} shows a run with the parameters: $\bar{\sigma}=0.1$, $dt=1\times\bar{\sigma}$, $T_{{\rm max}}=1\times L$, $L=W=10$ and a chosen randomness of 75\% with 75 152 randomly generated events, actual randomness, 75,152\%, 10 families, 100 000 events. We can clearly see no structure arising even after 100 000 events, so no need to go higher than 75\% randomness. We also get a convincingly even distribution of events among the families considering that $T_{{\rm max}}$ is small and we don't have 99.9\% randomness, so this limit checks out.\newline	
		 The other three panels are to check the deterministic limit. Plot \textbf{(b)} has the following parameters: $\bar{\sigma}=0.1$, $dt=1\times\bar{\sigma}$, $T_{{\rm max}}=1\times L$, $L=W=10$ and a chosen randomness of 0.01\% with 0 randomly generated event, actual randomness, 0\%, 10 families, 10 000 events. Plot \textbf{(c)} is a run where we just increased the number of events to 100 000 with 13 randomly generated events, actual randomness 0.013\%. Plot \textbf{(d)} is just a close up of the bottom left plot so the phase transition is more clear.\newline
	Plot \textbf{(b)} is exactly what one would obtain from a run of the deterministic code. However there were no actual events being created randomly. Plot \textbf{(c)} shows a clear breaking of the crystal with five randomly generated events happening exactly around 380 000 Minkowski time units. This shows that randomly generated events are responsible for breaking the crystal and how sensitive the system is to randomness and for this reason we can't really say that even with very small randomness (like 0.000001\%) we get exactly the results of the deterministic case where the crystal structure, once fully formed, never gets broken. It seems to be the case that one randomly generated event might be enough to force the system out of the ordered phase. This may also suggest we are injecting very strong form of randomness into the system which might have to be relaxed in the future.}
	\label{3runsfinal}
\end{figure}

Let us now see if we get the the other two programs as limits of this one starting with the random limit. Figure~\ref{3runsfinal} shows three runs, the first (plot on the top) being the random limit 75\% of randomness and we can clearly see that no structure forms. We got the following distribution of events:
\begin{multicols}{2}
	\begin{itemize}
		\centering
		\item Family 1: 8.3\%
		\item Family 2: 8.3\%
		\item Family 3: 8.0\%
		\item Family 4: 8.4\%
		\item Family 5: 13.6\%
		\columnbreak
		\item Family 6: 9.3\%
		\item Family 7: 7.6\%
		\item Family 8: 12.0\%
		\item Family 9: 11.9\%
		\item Family 10: 12.6\%
	\end{itemize}
\end{multicols}

We could go further and increase the probability even more, but there is no need, the main thing to check is that even in a long run, no structure arise and considering we didn't went with like 99\% randomness and, remember, a bigger $T_{{\rm max}}$, the distribution of events we obtained is more than convincing. 

Now for the deterministic case, we already said that the system is highly sensitive to the random input we build into the dynamics and it seems to be the case that one random event might be enough to break the crystal and force the system into a disordered phase again. The last plots of Figure~\ref{3runsfinal} show two runs with 10 pasts, 10 000 and 100 000 events respectively. In the plot on top right , with 10 000 events, a crystal forms and doesn't break, however no random event was actually generated. So this should be exactly like the deterministic case since the code is the same. But take a look to the bottom plots, the one on the right being a close up of the run with 100 000 events in the left. There were 13 randomly generated event and the last five randomly generated events of the run were events number 76 471, 82 442, 88 069, 91 965, 99 553 and their collision time were respectively, 375 948, 388 642, 386 908, 388 637, 388 643. The other 8 happened before the formation of the first crystal and they played their part too.  We will discuss these results in the next chapter but note how the collision times of these only five events coincide with the breaking of the fully formed crystal...

\section{Discussions of the Results}

In this section we will discuss our results. There is still information missing to a complete and final interpretation, that is for future work which will be discussed in the next chapter. The objective of this Thesis is to obtain the depicted phase transitions in chapter 4 and to control it the best we can with our parameters.

\subsection{Deterministic results}

\begin{figure}[htbp]
	\centering
	\includegraphics[width=120mm]{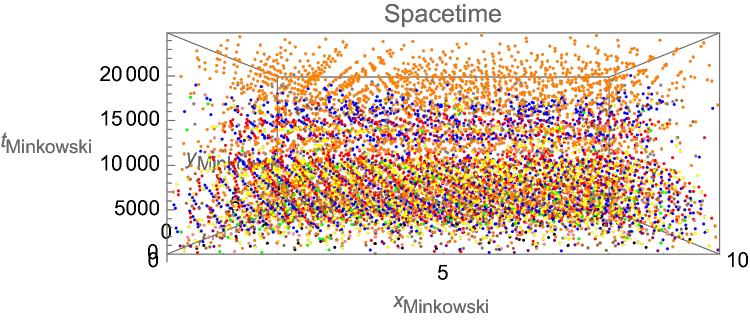}
	\includegraphics[width=120mm]{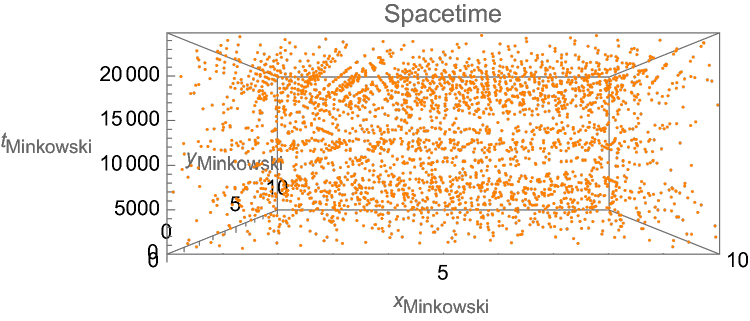}
	\includegraphics[width=120mm]{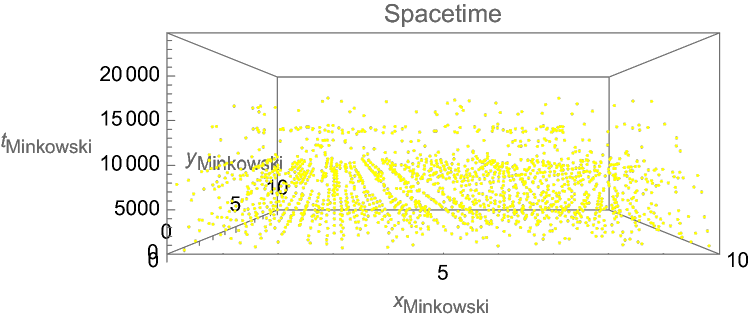}
	\includegraphics[width=120mm]{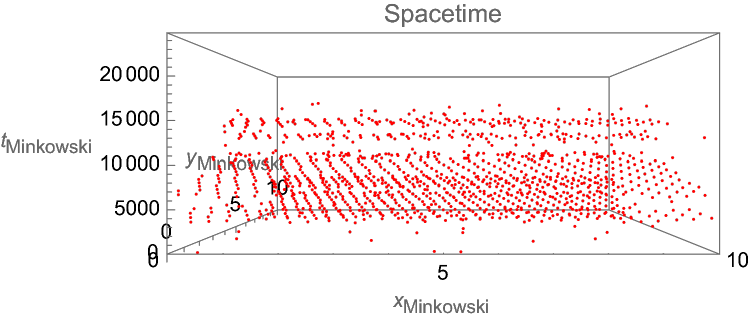}
	\caption{A run of the deterministic program showing different families trying to form the crystals simultaneously. Parameters: $\bar{\sigma}=0.1$, $dt=1\times \bar{\sigma}$, $T_{{\rm max}}=1 \times L$, $L=W=10$, 10 Families, 10 000 Events. The top plot has all the families, with the ones below showing individual families.}
	\label{determdifferfam} 
\end{figure}

Something interesting to note is the time lapsed in each run. It seems, and this happens in every run, that the quicker the crystal emerges, the longer the run goes for, in terms of Minkowski time lapsed. Naturally $T_{{\rm max}}$ will also affect this since if it is bigger we are allowing the particles to go through more windings before interact. But $T_{{\rm max}}$ aside there is still a difference. So basically when the crystal is formed Minkowski time seems to be going to the future at a faster rate per event created. This is a hint to how the structures are forming. We believe that this might be due to a loss of discausality when we are in the crystal phase. In other words, we are inclined to think that when the system enters this phase the total and partial orders get aligned so Minkowski time keeps going forward with each new event. Also worth noting is that there seems to be a dominance of one family after the structure is formed. This isn't always the case, sometimes families fight over each other until one wins as figure~\ref{determdifferfam} shows. In the end it seemed that the orange family was about to win. One can say this because the top plot shows no more disorder in the end of the run, there are not events from other families, so, in effect, one can say that the crystal was fully formed. 

\begin{figure}[t!]	
	\centering
	\includegraphics[width=150mm]{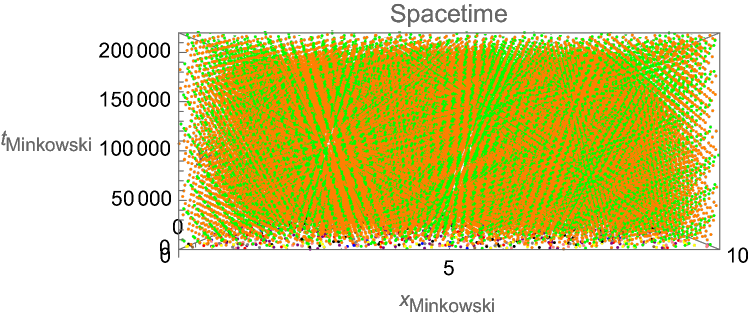}
	\includegraphics[width=150mm]{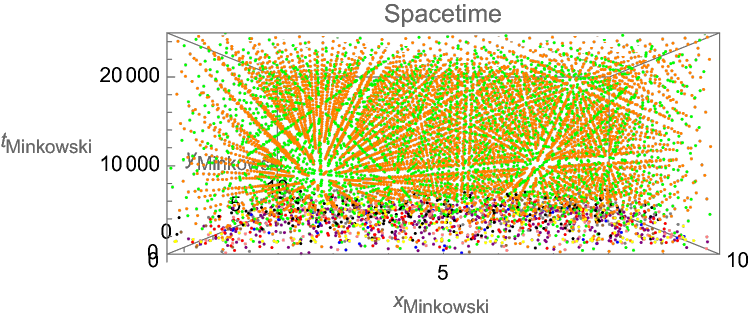}
	\caption{A run of the deterministic program showing two families completely forming the crystal. Parameters: $\bar{\sigma}=0.1$, $dt=1\times \bar{\sigma}$, $T_{{\rm max}}=1 \times L$, $L=W=10$, 10 Families, 100 000 Events. The lower plot is  a close up to better see the phase transition. This run confirms that more than one family can fully form a crystal, which is a good indicator again that the number of different families remains conserved through out the run.}
	\label{det2fam} 
\end{figure}

Now, we already discussed and saw, so we know for sure, that the families are preserved through out the run, no family disappears from the thick present. So why only one family form the crystal? After all, in the 1d case yes, some families stop interacting after a while, but there is usually more than one giving rise to quasi-particles. Well here this can happen to as figure~\ref{det2fam} shows, however it is much more uncommon. We think this is due to the initial conditions again. In one dimension the initial positions don't really affect how easily or not two particles will interact, any two particles always meet within one winding. In two 2d things get much more complex and it seems that initial conditions play a much more important role in determining the emergence of quasi-particles, in particular, favouring some families over others to form the crystals. 

We actually think these two\footnote{The effect described on the previous paragraph and the one we are describing now.} phenomena could be related. It seems to be the case that discausality can play a role in the competition between families, breaking crystals that were about to form, just like randomness. So why doesn't it break the crystal after it forms? Because, like we said, it seems that discausality stops after that which explains why Minkowski time goes forward more quickly. Going a bit further with our hypothesis, since there is no random input in the code and the fact that process by which the parents are selected is fully deterministic, eventually time will dominate and the closest pasts will always be between an event and their children, which explains why there is no more discausality and hence time goes forward at a faster rate per event created. This seems like a plausible explanation, but there is not enough information yet to conclude this with certainty, we will talk about this a bit more when we talk about how we can study better quasi-particles.

Let's talk about the thickness of our particles now. We saw that if the radius got bigger than 20\% of $L$ no structure was formed. This is because most of the particles have a great probability of being overlapped. So what happens is that, the first entry in the ``Distances" table will always be negative for most pairs. So the code will have to look a lot until it finds one pair that isn't already overlapped. The fact that this is constantly happening makes our dynamical rule of selecting the closest pasts obsolete. So it's almost like picking our events randomly. If you go over $\bar{\sigma}=0.5\%$ then the run will most probably be aborted because all the particles and anti-particles have a high probability of being all overlapped and no pair will be found. This is naturally an issue with the code. This is not something to worry about like we already said, since we don't want particles to take so much space, in fact, we want them as small as possible.

We will come back to some of these issues in chapter 6 since they are more complex than they seem and some of them lie at the heart of the foundational problems of quantum gravity. Since the discussion of the results of the random code corresponded to their mere presentation we now move one to the discussion of the results of the final code.

\subsection{Results of mixing randomness and determinism}

There will be a lot to say here in the future, for now we will focus on the results and a possible explanation. Again, and we can't stress this enough, the objective of the Thesis was to obtain the complex phase transition between disorder and order. So that is what we will be focusing in this small subchapter. We will be discussing figure~\ref{runfinalmore} and the last plot in figure~\ref{3runsfinal} (the close up). We have already said the main things on the discussion of the deterministic results, these are just some final remarks on a really important and difficult topic. 

We have already said that it seems that one randomly generated event might be enough to break the crystal, here is a possible explanation. When a randomly generated event happens, discausality might take place. That means an event created later will be embedded in the manifold causal past of some of its ancestors\footnote{Never the parents though!} in the total causal order\footnote{The order by which events are created.}. That might be enough to trigger interactions between different families that stopped interacting and whose time coordinates are small when compared with the events in the thick present. However, if a new event is created with a Minkowski time of the order of one of those families, it might just be enough to make one of them to have the closest past to the family of that new event therefore making that family start to interact again. This can trigger a cascade phenomenon making other families that also stopped interacting start to interact again for the same reason. 

Figure~\ref{runfinalmore} shows clearly families fully forming a crystal followed by random phases. But figure~\ref{3runsfinal} gives us the biggest hint because we know which events were created randomly and when. Take a look at the end of section 4.3, those last 5 randomly generated events all have a time of birth right around the time the crystal was broken. More even, the third event, event number 88 069 was embedded in the past of events 76 471 and 82 442, a clear example of discausality. This again reinforces our idea of how the quasi-particles might be forming. Finally, and due to what we discussed, it seems that one randomly generated event might be enough to break break the crystal, so it is not clear yet if this code has a deterministic limit in the sense that once the crystal forms the system never leaves that phase, or at least after a certain point that happens. On the other hand this might be suggesting that the way we are introducing randomness into the dynamics is too strong and we might need to relax it a bit.

\subsection{2d quasi-particles}

We already presented a hypothesis as to how the crystals might be forming. We don't know for sure yet because we don't have enough information. What is missing is the causal structure in the set so we can see clearly the causal connections and what the system is doing. In the Minkowski manifold depicted here in the figures we do not see the causal structure nor the causal network underpinning the emergence of flat spacetime originated by the causal irreversible evolution rule. In order to understand an event's lineage as well as its birth order one would have to analyse the causal sequence supporting the Minkowski embedding. This causal sequence cannot be superposed on the graphic while depicting the Minkowski spacetime. A full treatment of the causal network is beyond the scope of this Thesis and is left for the follow-up analysis by our collaboration's upcoming work.

We can say however that if in fact it can take only one randomly generated event to break a limit cycle, and the results suggest that, then this might mean the way we are introducing randomness into the dynamics is too strong. There are several hypotheses to try in the future like trying to replicate the way randomness was introduced in the 1d case by choosing randomly $T_{{\rm max}}$ of out a finite set of possibilities for each event. Before doing that though, we need to understand exactly how the structures are forming and if, in fact, one random event is enough to break them. 

On the topic of randomness, it seems that adding an extra more dimension makes the initial random input much stronger than expected. This is something worth studying as well, since we plan on a 3+1d version of this work.

\section{Defining an Intersection}

We are going to discuss the definition of an intersection in this section. We would like to start by saying that in order to reach our goal we made as many simplifications as needed to have a simple treatment of collisions. That being said, we are naturally not computing the intersections in a particle physics scattering. Neither is the goal to compute the particle collisions in a particle physics model, but rather it is to observe the already discussed phase transition. We want to define an interaction to be not too cumbersome operationally and computationally, because we are interested in the overarching effect generated by half a million collisions which gives rise to phase transitions in system dynamics and gives emergence to structures like those observed in the 1+1d quantum gravity model. 

To that end, for our purposes, we are interested in simplifications of individual collisions that if used consistently will give the desired global behaviour. We tried many ways of defining the collisions and none seemed to work for our purpose. In fact the emergence of quasi-particles was something we obtained only after much effort and after realising something about the dynamics that had been missing to us this whole time and which we're about to explain. 
\subsection{Derivation of collision coordinates}

We talked about how we compute the intersections but we didn't specify what coordinates we give to the new event. This is definitely not an easy task. As we discussed in chapter two, the initial conditions fully determine every possible collision between any two pairs in the 1d case, determining the full finite and constant set of spatial positions for all future interactions. Here, the same should happen. Note also that is thanks to this that quasi-particles emerge. The thing is that in the 1d case defining coordinates for the new event is simple, you just take the coordinates of the lines representing the particles in spacetime. But now that particles have an extent, events are no longer points but extended regions in spacetime. Yet we need to make collisions points in spacetime. What point should we define to be the coordinates for the new event? One could take, analogously to the 1d case, the intersection of the centers of the particles. But firstly there are events where particles might just graze each other, and secondly the probability that this will happen is 0. 

\begin{figure}[t!]
	\centering
	\includegraphics[width=150mm]{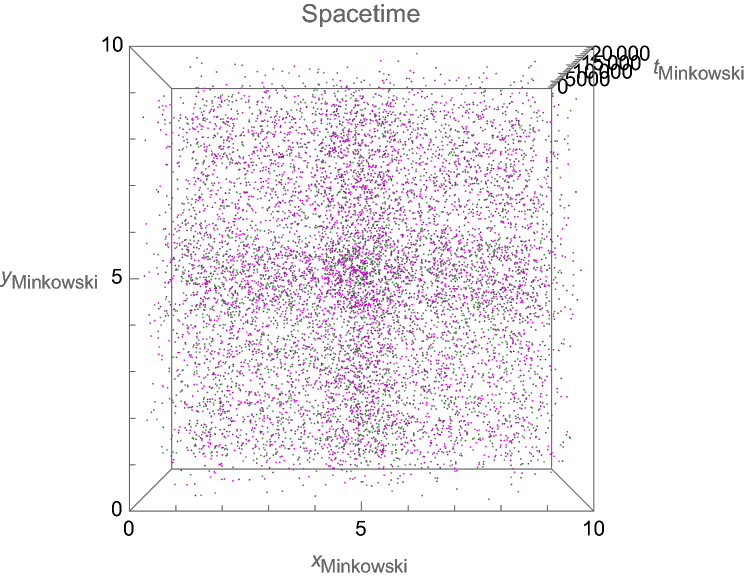}
	\caption{A top view of a run from the fully random program to show the effect of choosing the coordinates of the events to be in the middle of the centers of both particles. Parameters: $\bar{\sigma}=0.1$, $dt=1\times \bar{\sigma}$, $T_{{\rm max}}=1 \times L$, $L=W=10$, 5 families and 10 000 events. We choose the random program to better highlight this effect, since no structures arise here.}
	\label{topview} 
\end{figure}

The most natural candidate is the midpoint between the centers of the particles, but this raises two problems. First when you do this, care must be taken in that the particles may interact at the boundary and their centers are on opposite sides of the boundary, for instance one at $x=9.9$ and the other at $x=0.1$. If we take the mid-point between these two, it will be $x=5$. This is clearly wrong and the effect is the cross you can see in figure~\ref{topview} obtained from the fully random program. We chose this program for this because we don't want to other structures to show. In fact, in this case, you should define the distance to be the one through the boundary. So if the distance\footnote{We are referring to the distances in both $x-$ and $y-$axes.} is smaller than half the side of the square the position of the new event is just the average, if its bigger then we need to take their sum and use the ``Mod'' function to bring the particle back to the square. 

But this didn't seem to solve the bigger issue. In fact, if at each new event we redefine the position of the rays that gave birth to it, by shifting them to a mid point, the effect is like changing the initial position of that ray, therefore adding new coordinates to the initial conditions. Bear in mind that the initial conditions determine all future possible spatial positions for collisions. In the 1+1d case it is the fact that there is a constant number of initial conditions that allows for some of the spatial position to be exhausted and the phase transition to occur in the form of emergent quasi-particles. In this 2+1d model shifting the collision coordinates to a midpoint is the equivalent of constantly adding new initial conditions. When we create a new event we are in practice restarting the run all over again. The result is that the phase transition never occurs. The non-deterministic part of the algorithm is never overcome. This was the case in our work for for a long period.

\subsection{Solving for event coordinates: simplification}

Our solution was to define the coordinates of the new event to be the ones of the particle. This naturally means that the anti-particles will be shifted at each interaction\footnote{Remember that both particle and anti-particle are supposed to leave the same point, in this case the position of the new event.}. Shouldn't this be a problem? Well, since that for each particle there is an anti-particle that shares the same initial conditions, we only need to not alter one of them and keep choosing consistently in the future which coordinates to change. We tried many, many things but nothing seemed to work until we realised this. So that is why we choose, for consistency the coordinates of the final parent. But then we asked, could we choose the particle to be the final parent and the coordinates for the new event to be the ones of the anti-particle? No, we do that, no structure emerges. Why? Again we don't know for sure yet but we suspect that is because the number of different families in the table of the anti-particles, ``livev2'', is not conserved since they are not the final parent. Again we need the causal structure to understand this better. This is however another hint at how the structures are forming.

We can come up with different strategies in the future to try and solve this, either by changing the dynamics and therefore how we define an interaction or by trying some more esoteric ideas like trying to create a grid of spatial positions which would work great for the quantum regime which is also intended for pursue in future work. The problem is that if we discretize the space we will also have to discretize time, but then what does velocity mean, what is velocity for that matter? Velocity is a continuous concept that doesn't even make since without continuity. The variable $dt$ is already proportional to the thickness/radius since it determines the precision with which we compute an interaction. But for now it is just a way of saying how much precision we want in terms of the thickness. It's an easier and more concrete way of talking about the time precision we want, however, were we to discretize time, this would be a theoretical necessity. These are all deep problems for which we have no answer yet and lie at the heart of the foundations of a theory of quantum gravity.

\section{Step-by-step overview of results}

\begin{enumerate}
    \item Derivation of parametric equations for the spacetime intersection of momentum-carrying null rays.
    
    \item Derivation of the rules under which rays represented by point particles might interact, adopting a modified version of a particle physics' cross-section.
    
    \item Development and implementation of an algorithm describing the random interaction between null rays, when the closest pasts rule is not applied. This step is essential to test the robustness of the algorithm, before a rule for the interacting rays is applied. The homogeneous distribution of events amongst families that was obtained reveals the robustness of the code, and that, indeed, no consistent algorithm was used in the random interaction case.
    
    \item Adaptation of the previous algorithm to consistently reflect the selective interaction of the null rays of similar lineage in the causal network, we call this the `closest pasts rule'.
    
    \item Obtaining of the phase transition into the time-symmetric regime, which is key to the success of the ECS program as a proposal for quantum gravity. This phase transition is marked by the emergence of lattice-like structures, very much like crystals.
    
    \item Robustness test of the closest pasts code in both the fully deterministic regime, and in the fully random regime. Both of these regimes proved to be sound and as expected according to the dynamics of the algorithm.
    
    \item Identification of limit cycle behaviour in the crystal structures obtained by phase transitions. This goal of describing 2+1d ECS in the language of basins of attractions of discrete dynamical systems was initially beyond the expectations of this thesis. In the 1+1d case the limit cycle behaviour was discovered only years after the seminal work. The fact that limit cycles have been observed in the 2+1d case is an extra test of robustness that the simulations here developed are a fiducial match to the initial models formulation, reflecting an arrow of time that leads to symmetric evolution.
    
    \item Carrying out several robustness tests for experimenting on the algorithm, namely with:
    
    \begin{enumerate}
        \item large number of families, in the form of additional randomness of the initial conditions,
        \item longer runs increasing the total number of events to enable the comparison with
        \item different mixes of determinism and non-determinism, to check if the observed effects are indeed according to what would be expected.
    \end{enumerate}   
    
    \item Itemization of items, which are subject of future work, exhaustively listed and will be part of our upcoming study. 
    
    \item Lastly the realisation was made that a further extension and completion of ECS to the 3+1d case might be fairly straightforward, and not involve further cumbersome puzzles as those encountered in the current project.
\end{enumerate}

\section{Conclusion and future work}

Quantum gravity has plagued the minds of theoretical physicists for the past several decades. Despite a large community attempting to find a gravitational theory for the quantum regime, there is no one attempt that remains unchallenged as a successful solution \cite{carlo, 3roads}.

The lack of a satisfactory answer thus far might suggest that the discord between the classical and quantum regimes lies deep within the fabric of modern physics. In 2013 Cort\^{e}s and Smolin proposed a novel approach to the problem developed from a radically different principle: the premise that the fundamental theory of reality breaks time-reversal invariance in the quantum gravity regime \cite{paper1}. 

The second premise in their work is that, together with the passing of time (encoded as a principle of causality), energy-momentum are fundamental variables and not emergent with the embedding of the causal network. These two premises distinguish the ECS proposal from Rafael Sorkin's very promising program of causal sets \cite{causalsets, csfd, csfd1, csfaydowker1, csfaydowker, csfaydowker2}.
	
Cort\^{e}s and Smolin's initial work gave origin to a succession of studies testing their radical principle in novel areas, namely,
\begin{itemize}
\item time-asymmetric extensions of general relativity, and the development of observational signatures \cite{TA1, TA2};

\item establishing a correspondence between their program and spin foam models of lattice gauge theories \cite{spinfoams};

\item recognising the dynamics of ECS as an example of the one found in discrete dynamical systems described by \cite{AW}, and consistently identifying the limit cycles of each model \cite{limitcycle}   

\item they then proceeded to examine whether the causal structure that underpins the emergence of flat spacetime manifold might exhibit a form of violations of causality in the quantum regime. This is named retrocasality and was first identified in the two state vector formalism by Yakir Aharanov and collaborators \cite{tsvf}. In work together with Aharanov's collaborator Avshalom Elitzur and Eliahu Cohen they concluded that there is no violation of causality in ECS like the two state formalism exhibits. Instead they observed that the order of generation of individual events in the causal network might be opposed to the order of Minkowski time of the embedding manifold. This was called disordered causality \cite{paper3, paper4}.

\item More recently the same questions raised by a fundamental arrow of time have made possible for progress to be made in the areas of complex systems within a theoretical physics formalism, more precisely in the biology field, where a theoretical physics description of living systems has been attempted. Cort\^{e}s, Kauffman, Liddle and Smolin have named their proposal ``Biocosmology'' \cite{biocosm1,biocosm2,biocosm3}.

\end{itemize}

The seminal work in 2013 treated exclusively the 1+1d regime and showed a phase transition between time asymmetric dynamics and a time symmetric regime, by means of computer-run simulations. Since then the question remained within the collaboration of whether those results would be observed in higher dimensional spacetimes. 

A successful observation of the phase transition phenomenon in 2+1d and 3+1d manifolds is a crucial step towards the development of a consistent quantum gravity proposal. Other features like curved space-time and cosmological expansion are formally included in the seminal paper. The inclusion of matter in the emergent spacetime is also possible to address since the events are endowed with energy-momentum by construction. 

The considerable challenge of the past decade in the ECS program has remained the higher dimensional extension. The main obstacle lies in the fact that the intersection of null lines carrying energy-momentum in a formalism where the lines are treated as real numbers has a set of solutions of measure zero. This is not the case in 1+1d where solutions always exist. 

It is therefore absolutely imperative to draw attention to the fact that, at the start of the present Master's program, there was no reason to believe that this obstacle could be overcame. It was in fact very likely that the zero-measure set of solutions would form a stumbling block and no interaction of events could ever be described, thereby hindering progress in no uncertain terms.  
It is therefore of no surprise that the results obtained in this work have surpassed all expectations. 

We were able to develop an algorithm to describe interactions between null lines given random initial conditions in the two dimensional spatial plane. Then we designed an algorithm that solves parametric equations and finds when any given two given null lines endowed with energy-momentum will intersect in the future within the assumption of periodic boundary conditions imported from the 1+1d case. 

In the simulations of energetic causal sets we are interested in the dynamics of evolution of systems given adjustable amounts of time symmetry and time asymmetry. This explains the simplifications adopted when describing null-ray interactions. More specifically we are studying whether a phase transitions does indeed occur between the time asymmetric to the time symmetric regime, as reflected in the 1+1d case. In the 2+1d case we have observed that a crystal emerges which corresponds to the quasi-particle trajectories described in the award-winning work.\footnote{Ref.~\cite{paper1} was awarded First Place in the Inaugural Buchalter Cosmology Prize announced at the $225^{th}$ meeting of the American Astronomical Society in Seattle, WA, on January $6^{th}$, 2015 \cite{buchalter}.}

After the emergence of the crystal, i.e.\ time-symmetric phase, we can then proceed to study how does this emergence depend on the relative percentages of time symmetry and time-asymmetry in the dynamics. For our purposes of studying model behaviour we can think of the ratios of time symmetry to time asymmetry in the dynamics, as ratios of deterministic versus random evolution, which are a property of the algorithm specified by parameters which can be tuned for different ratios. This was tested for several amounts of deterministic versus random evolution steps. 

It is rather curious that in 2+1d the amount of randomness input to the system at the initial time--in the form of the initial coordinates and initial velocity of each event (typically of order 5 to 10 initial events) has a much greater influence on the percentage of deterministic versus indeterministic behaviour of the system, than in the 1+1d case. 

The total amount of steps spent by the system evolving out of the time irreversible phase (disordered structure) and crossing over to the time reversible phase (crystal) is here much larger than observed in the 1+1d. This rather interesting detail merits substantial attention in our follow-up work, that will enable us to understand what parts of the evolution do, in actuality, determine the overall amount of randomness present in the system at any one time.
As an unexpected bonus we observed the emergence of distinct limit cycles as in the work \cite{limitcycle} which identifies ECS with the dynamics of discrete dynamical systems \cite{AW}. Discrete dynamical systems, which we described in Section~\ref{1dsim}.
 This evolution towards limit cycles in the 2+1d case matches that observed in the 1+1d case rather closely. 
 
Just as in the 1+1d case, the limit cycle approached is unstable in that, due to the amount of randomness present in the system, the late-time evolution is characterised by oscillations in and out of limit cycles.

What is more, here in the 2+1d case, the sensitivity to small percentages of randomness is such that even only one event, chosen outside of the deterministic rule, is enough to destabilise the limit cycle; throw evolution back in the time asymmetric (disordered) regime; only to be caught again by a limit cycle once enough events have taken place in the deterministic regime.

Note that we have not yet performed an exhaustive analysis of the causal network underpinning the emergence of Minkowski spacetime. A complete sequencing of the causal and birth orders will reveal the dynamics that play under the space-time structure. At that point it shall become apparent the forms of disordered causality that the system might reveal and cast light on the evolution of the quantum counterpart of the 2+1d ECS. 

Finally, while taking the problem from one to two spatial dimensions has forced us to overcome substantial and cumbersome problems, i.e.\ in the adoption of new points of principle (leading to introduction of a particle cross-section, $\bar{\sigma}$, and a particle/antiparticle quantum number), we envisage that there will be no new obstacles in further extending to three spatial dimensions other than increased computational demand. It appears that the final 3+1d completion of the emergence of spacetime might not be very far off.

All of this in-depth analysis is the focus of our upcoming work. For now, it is revitalising to observe a consistency of the behaviour of fundamental time irreversibility models of quantum gravity in their evolution from 1 to 2 spatial dimensions.

The energetic causal set program as a model for quantum gravity starts with a very small change in what the evolution of the physical variable `time' represents. Such a subtle change of explicitly incorporating time irreversibility in the equations describing reality has already borne more fruit than the authors could ever have ambitioned it to.


\clearpage
\end{document}